\documentclass[11pt,letterpaper]{article}
\usepackage{jheppub}
\usepackage{pstricks}
\usepackage{skak}

\graphicspath{{./Figures/}}

\newcommand{\beq}{\begin{equation}}
\newcommand{\eeq}{\end{equation}}
\newcommand{\bea}{\begin{eqnarray}}
\newcommand{\eea}{\end{eqnarray}}
\newcommand{\be}{\begin{eqnarray}}
\newcommand{\ee}{\end{eqnarray}}
\newcommand{\C}{\mathbb{C}}

\newcommand{\R}{\mathbb{R}}

\def\tilde{\widetilde}
\def\hat{\widehat}
\def\bar{\overline}

\def\CG{{\mathcal G}}

\def\CN{{\mathcal N}}
\def\CO{{\mathcal O}}
\def\CP{{\mathcal P}}

\def\CS{{\mathcal S}}
\def\CT{{\mathcal T}}

\def\CW{{\mathcal W}}

\def\QQ{\mathcal{Q}}
\def\GG{\mathcal{G}}
\def\II{{\mathcal I}}

\newcommand{\te}{e}
\newcommand{\bbC}{{\mathbb C}}
\newcommand{\bbCP}{{\mathbb C \mathbb P}}
\newcommand{\bbL}{{\mathbb L}}
\newcommand{\ttL}{{\mathtt L}}
\newcommand{\ttR}{{\mathtt R}}
\newcommand{\cT}{{\mathcal T}_{4d}}
\newcommand{\Tr}{{\rm Tr}}

\newcommand{\nupr}{{\nu^\prime}}

\newcommand{\TS}{\CT_{2d}}

\newcommand{\fq}{{\mathfrak q}}
\newcommand{\fp}{{\mathfrak p}}
\newcommand{\ft}{{\mathfrak t}}

\providecommand{\tabularnewline}{\\}
\setboardfontsize{8}

\title{$2d$ Index and Surface operators}

\preprint{CALT-68.2932}

\author{Abhijit Gadde}
\affiliation{California Institute of Technology \\ Pasadena, CA 91125, USA}
\emailAdd{abhijit@caltech.edu}

\author{and Sergei Gukov}
\emailAdd{gukov@theory.caltech.edu}

\abstract{
In this paper we  compute the superconformal index of $2d$ $(2,2)$ supersymmetric gauge theories. The $2d$ superconformal index, a.k.a. \emph{flavored} elliptic genus, is computed by a unitary matrix integral much like the matrix integral that computes $4d$ superconformal index. We compute the $2d$ index explicitly for a number of examples. In the case of abelian gauge theories we see that the index is invariant under flop transition and CY-LG correspondence. The index also provides a powerful check of the Seiberg-type duality for non-abelian gauge theories discovered by Hori and Tong. 

In the later half of the paper, we study half-BPS surface operators in $\CN=2$ superconformal gauge theories. They are engineered by coupling the $2d$ $(2,2)$ supersymmetric gauge theory living on the support of the surface operator to the $4d$ $\CN=2$ theory, so that different realizations of the same surface operator with a given Levi type are related by a 2d analogue of the Seiberg duality. The index of this coupled system is computed by using the tools developed in the first half of the paper. The superconformal index in the presence of surface defect is expected  to be invariant under generalized S-duality. We demonstrate that it is indeed the case. In doing so the Seiberg-type duality of the $2d$ theory plays an important role. 
}


\begin{document}
\maketitle

\section{Introduction}

In recent years, a lot of progress has been made on exact results in supersymmetric gauge theories. A quantity that is especially amenable to exact computations is the superconformal index. It can be defined for any superconformal gauge theory as the super-trace over the Hilbert space in radial quantization. As is the case with any Witten index, the superconformal index is protected against quantum corrections and deformations that preserve the superconformal symmetry. As a result, the superconformal index can be computed in the free limit in case SCFT admits such a limit. Despite being easy to compute, the superconformal index encodes very useful physical information. It counts all the short multiplets of the theory modulo the equivalence relation that sets to the short multiplets those can recombine to form a long multiplet \emph{c.f} \cite{Gadde:2009dj}. The superconformal index, or index for short, was computed for four dimensional gauge theories in \cite{Romelsberger:2005eg,Romelsberger:2007ec,Kinney:2005ej}. In \cite{Dolan:2008qi,Spiridonov:2008zr,Spiridonov:2009za,Gadde:2010en} it was used to probe four dimensional Seiberg/toric dualities. It was also used in \cite{Gadde:2009kb,Gadde:2010te,Gadde:2011ik,Gadde:2011uv,Gaiotto:2012xa,Beem:2012yn} to test newly conjectured dualities for the ${\cal N}=2$ and ${\cal N}=1$ theories of class $\cal S$. The three dimensional index has also proved useful in analyzing $3d$ ${\cal N}=2$ theories of class ${\cal R}$ \cite{Dimofte:2011py}. In this paper, we take the next natural step and compute the superconformal index for $2d$ $(2,2)$ gauge theories.

Of course, a quantity closely related to the $2d$ superconformal index, the elliptic genus, of $(2,2)$ theories has been widely studied in the literature. Our treatment of the $2d$ index differs from that of the elliptic genus. The elliptic genus was discovered in \cite{Schellekens:1986yi,Schellekens:1986xh,Witten:1986bf} and has since been computed\footnote{The references cited are but a few examples. They are not meant to be exhaustive.} for Landau-Ginzburg models \cite{Witten:1993jg,Berglund:1993fj,Berglund:1994zg}, WZW/coset models \cite{Henningson:1993nr,Troost:2010ud,Ashok:2011cy,Ashok:2012qy,Eguchi:2004yi,Eguchi:2006tu} and supersymmetric non-linear sigma models on Calabi-Yau targets \cite{Eguchi:2004yi,Eguchi:2008ct}. In almost all the computations, the modular property of the elliptic genus plays a crucial role, see  \cite{Gaberdiel:2008xb} for the application of modularity. In this paper, we approach the problem from the point of view of the gauge theory and derive a matrix model to compute the index. It is well known that a  supersymmetric $2d$ gauge theory leads to the non-linear sigma model on its moduli space. For non-anomalous gauge theories without superpotential, this moduli space is a non-compact Calabi-Yau manifold. The flavor symmetry of the gauge theory maps to the Calabi-Yau isometry. We refine the $2d$ index by equivariant parameters a.k.a. flavor fugacities $a_i$ that couple to the Cartan subgroup of this symmetry. As a result, our $2d$ index is a function of $q,y$ and $a_i$ as opposed to the conventional elliptic genus that is function of only $q$ and $y$. The target space can be made into a compact Calabi-Yau by introducing an appropriate superpotential. As we will see in the bulk of the paper, the elliptic genus of the resulting compact space can be easily obtained from the superconformal index or \emph{equivariant}/\emph{flavored} elliptic genus of the parent non-compact space.

In the second half of the paper, the tools developed in the first half will be applied to study the half-BPS surface operators in $\CN=2$ superconformal gauge theories. The half-BPS surface operators in $\CN=4$ super Yang-Mills have been studied in great detail in \cite{Gukov:2006jk,Gukov:2008sn}. The authors define the surface operator by prescribing a co-dimension two singularity in the field configuration that preserves half the supersymmetry. They also show that such singularity can be induced in the field configuration by first coupling the $4d$ theory to new degrees of freedom living on the support of the surface operator and later integrating them out. The two dimensional theory can have description in terms of a $(4,4)$ supersymmetric non-linear sigma model or a $(4,4)$ supersymmetric gauge theory. It is the latter description that will be of most use for our purposes. Compared to $\CN=4$, the surface operators in $\CN=2$ theories remain somewhat unexplored. For the theories of class $\cal S$, they are related to degenerate vertex operators of Liouville theory \cite{Alday:2009fs}. One can construct a subset of possible half-BPS surface operators by taking infinite tension limit of dynamical BPS vortex strings. The superconformal index of the surface operators of this type in $\CN=2$ theories of class $\cal S$ was computed in \cite{Gaiotto:2012xa} via indirect means. For a discussion on nonabelian semi-local vortex strings, see \emph{e.g.} \cite{Shifman:2006kd}. Corresponding brane construction has been described in \cite{Hanany:2003hp}.

In this paper, we utilize the construction of the surface operator as the $4d$ gauge theory coupled to $2d$ gauge theory. As long as the $2d$ gauge theory satisfies a certain criteria including $(2,2)$ supersymmetry, this construction engineers a half-BPS surface defect. We see that the subset of surface operators coming from the infinite tension limit of the vortex strings can also be obtained this way. We compute the superconformal index of surface operators in the UV where the $2d$-$4d$ system is weakly coupled. The $2d$ gauge theory index obtained in the first part of the paper will play an important role in this computation.

Outline of the rest of the paper is as follows. In section \ref{LGindex}, we will define the $2d$ index and compute it for the $(2,2)$ chiral and vector multiplets which are the basic building blocks of $(2,2)$ supersymmetric gauge theories. We will also compute the index of Landau-Ginzburg models and in the process learn an important lesson about the effect of superpotential terms on the index. In section \ref{abelian}, we compute the index of abelian gauge theories and study the flop transition and CY-LG correspondence in this context. Section \ref{nonabelian} is devoted to non-abelian gauge theories. We will perform a powerful check of the Hori-Tong duality by computing the index on either sides of the duality. In section \ref{physical}, we discuss the physical interpretation of $q$-difference equations satisfied by the $2d$ index. The remaining part of the paper is devoted to the study of surface operators. In section \ref{N4} we consider the surface operators in $\CN=4$ SYM. These surface operators are classified by so called Levi subgroups of the gauge group. We engineer the $2d$ gauge theory for surface operator of arbitrary Levi type. We then move to surface operators in $\CN=2$ theories in section \ref{N2}. We compute their index and demonstrate its invariance under the S-duality of the four dimensional theory. This computation involves identification of  the embedding of $2d$ $(2,2)$ superconformal algebra into $4d$ $\CN=2$ superconformal algebra. In the last section, section \ref{vortex}, we consider the special subset of surface operators obtained from the infinite tension limit of the vortex string and relate our index to the one obtained in \cite{Gaiotto:2012xa}. Some interesting properties of the multiplet indices are summarized in appendix \ref{properties}.

\section{$2d$ index}\label{LGindex}
Let us start by defining the $2d$ superconformal index for $(2,2)$ superconformal theories. We take it to be the flavored elliptic genus in the NSNS sector. The choice of NSNS sector over the RR sector will be justified in section \ref{N2} when we compute the index of four dimensional gauge theory in the presence of surface defect. The generators of $(2,2)$ superconformal algebra needed to define the superconformal index are: left-moving conformal dimension $H_\ttL$, left-moving R symmetry $J_\ttL$, left-moving supercharges $\CG_\ttL^{\pm}$ and their right-moving counterparts $H_\ttR$, $J_\ttR$ and $\CG_\ttR^{\pm}$ respectively. In addition, the $(2,2)$ theory can have flavor symmetry $F$ which commute with the entire $(2,2)$ superconformal algebra. We pick $\CG=\CG_\ttR^{+}$. Then the $2d$ index with respect to $\CG$ is:
\be\label{2dindexdef}
\II^{2d}(a_j;q,y)=\Tr (-1)^F q^{H_\ttL} y^{J_\ttL} \prod_j a_j ^{f_j}.
\ee
Here $f_j$ are Cartan generators of the flavor symmetry. The symmetries appearing in this definition, $H_\ttL, J_\ttL$ and $f_i$, have the property that they all commute with $\CG$. Through the standard arguments about the Witten index, only the states with $\delta= 0$ contribute to the index, where $\delta := \{\CG,\CG^\dagger\}=2H_\ttR+J_\ttR$.

One of the main objectives of the paper is to compute the superconformal index of the $(2,2)$ gauge theories. A general $(2,2)$ gauge theory can be constructed out of chiral multiplets and vector multiplets. One can also allow the matter fields to be in twisted chiral representation of the superalgebra but we will not be considering such representations in this paper. Using the argument for the protection of the superconformal index, it can be computed in the zero coupling limit of the gauge theory. In this limit, the role of the vector multiplet is essentially to impose the Gauss law. The index is computed by multiplying the contribution of all the multiplets and imposing the Gauss law by integrating over the gauge group. The integration prescription will be described in section \ref{abelian} and \ref{nonabelian}, for now, we compute the index contribution of individual multiplets.

\subsection*{Index of the chiral multiplet}
The chiral multiplet $\Phi$ satisfies ${\bar D}_{\pm}\Phi=0$. Its superfield expansion is given by,
\be
\Phi=\phi+\theta^\alpha \psi_\alpha+\theta^+\theta^- F,
\ee
where $F$ is an auxiliary field. In table \ref{2dletters} we list the component fields of the chiral multiplet with $\delta=0$ and their contribution to \eqref{2dindexdef}.  Note that a free chiral multiplet admits a $U(1)$ flavor symmetry $f$. We have introduced the fugacity $a$ for it.
Summing these contributions, we get the ``single letter index" $f_\Phi(a;q,y)$. The index of the chiral multiplet $\II_\Phi(a;q,y)$ is  given by its plethystic exponent $\mathtt{PE}$.
\begin{table}
\begin{centering}
\begin{tabular}{|r|r|r|r|r|}
\hline
 & $ H_\ttL $ & $J_\ttL $ & $f$ & index\tabularnewline
\hline
\hline
$\phi$ & $0$ & $0$ & $1$ & $a$\tabularnewline
\hline
${\bar \phi}$ & $0$ & $0$ & $-1$ & $1/a$\tabularnewline
\hline
$\psi_{+}$ & $\frac{1}{2}$ & $-1$ & $1$ & $-q^{\frac{1}{2}}a/y$\tabularnewline
\hline
${\bar \psi}_{+}$ & $\frac{1}{2}$ & $1$ & $-1$ & $-q^{\frac{1}{2}}y/a$\tabularnewline
\hline
\hline
$\partial$ & $1$ & $0$ & $0$ & $q$\tabularnewline
\hline
\hline
$\bar\psi_{-}^{(0)}$ & $0$ & $0$ & $-1$ & $-1/a$\tabularnewline
\hline
\end{tabular}
\par\end{centering}
\caption{\label{2dletters}Letters with $\delta=0$, the only letters contributing to the superconformal
index. The superscript of $\bar\psi_{-}^{(0)}$ indicates that only the zero mode of $\bar\psi_{-}$ contributes to the index.}
\end{table}
\be
f_\Phi(a;q,y)&=&\frac{a+1/a -q^{\frac12} a/y- q^{\frac12}y/a }{1-q}-1/a\nonumber\\
\II_\Phi(a;q,y)&=&\mathtt{PE}[f_\Phi(a;q,y)]:=\exp\Big(\sum_{n=1}^{\infty} \frac1n f_\Phi(a^n;q^n, y^n)\Big)=\frac{\theta(a q^{\frac12} /y;q)}{\theta(a;q)}.
\ee
We have defined $\theta(x;q):=(x;q)(q/x;q)$ and $(x;q)=\prod_{i=0}^{\infty}(1-xq^i)$. In what follows, we will be using the variable $t:=q^{\frac12}/y$ instead of $y$. With this redefinition, the index becomes,
\be\label{2dindexdef2}
\II^{2d}(a_j;q,t)=\mbox{Tr}(-1)^F q^{L_0+\frac12 J_0} t^{-L_0}\prod_j a_j^{f_j}.
\ee
As the index of the chiral multiplet will play an important role in the rest of the paper, we coin a new function for it
\be
\Delta(a;q,t):=\theta(at;q)/\theta(a;q)=\II_\Phi.
\ee
This function enjoys a number of interesting properties including modularity. They  are summarized in the appendix \ref{properties}.

Let us add the superpotential $W=\Phi^{k+2}$. This term breaks the $U(1)$ flavor symmetry $f$ and the theory flows to a fixed point where the superpotential is marginal. This is the famous Landau-Ginzburg description of $k$-th minimal model. In our notation, the r-charge of $W$ is $1$. This determines the r-charge of $\Phi$ to be $1/(k+2)$. In fact, the r-charges of all the states are obtained from r-charges of the corresponding states in the free theory via,
\be
r_{{\rm LG}}= r_{{\rm free}} +\frac{1}{k+2} f.
\ee
This dictionary allows us to compute the index of the interacting fixed point immediately $\II_{{\rm LG}}=\Delta((q/t)^{1/(k+2)};q,t)$. After the spectral flow to RR sector, the index exactly matches with the elliptic genus of of $k$-th minimal model \cite{Witten:1993jg,DiFrancesco:1993dg}. Note that for $k=0$, the superpotential is a mass term and the theory is empty at low energies. The index also trivializes to $1$ in agreement with this fact.  In summary, the superconformal index of a chiral field of r-charge $r$ is simply $\II_{\Phi_r}(q,t)=\Delta((q/t)^r;q,t)$. From the modular property \eqref{Deltamodular} it follows that,
\be
\II_{\Phi_r}(e^{-2\pi i/\tau},e^{-2\pi i\sigma/\tau})=e^{i\pi\frac{\sigma}{\tau}((1-2r)(\sigma-\tau)-1)} \II_{\Phi_r}(e^{2\pi i\tau},e^{2\pi i\sigma}).
\ee
Note that the coefficient $(1-2r)$ appearing in the exponent of the pre-factor is exactly the central charge $\hat c$  of the fixed point corresponding to the chiral multiplet of r-charge $r$. The superconformal index of a general $(2,2)$ theory also transforms covariantly under modular transformation. The central charge ${\hat c}\,$ is the coefficient of $i\pi \sigma^2/\tau$ in the exponent of the modular pre-factor.

\subsection{Index of Landau-Ginzburg models}
Now we consider a Landau-Ginzburg model with $N$ chiral fields $\Phi_i$. They interact with the superpotential $W(\Phi_i)$ which is pseudo-homogeneous:
\be\label{LGW}
\CW(\lambda^{r_{1}}\Phi_{1},\ldots,\lambda^{r_{N}}\Phi_{N})\;=\;\lambda\CW(\Phi_{1},\ldots,\Phi_{N}).
\ee
This implies that the r-charge of $\Phi_i$ in the IR is $r_i$. The superconformal index of the fixed points is then $\prod_i \Delta((q/t)^{r_i};q,t)$. On the other hand, the index of the free UV theory of $N$ chiral multiplet is $\prod_i \Delta(a_i;q,t)$, where $a_i$ is the fugacity for each $U(1)$ symmetry rotating each field $\Phi_i$ individually. Note that the index of the interacting theory is simply given by the substitution $a_i\to (q/t)^{r_i}$ in the index of the free theory. This exercise illustrates an important point that is much more universal:
\begin{itemize}
\item After adding the superpotential, a $2d$ theory flows to a non-trivial fixed point where the r-charges of operators are determined by imposing the marginality of superpotential.
\item The superconformal index of the IR fixed point is obtained from the index of free (or weakly coupled) UV theory simply by appropriate substitutions $a_i\to (q/t)^{r_i}$, assuming we have kept track of all the flavor symmetries present in the UV.
\end{itemize}
This observation leads to an immense simplification. It allows us to focus only on $(2,2)$ theories without superpotential. That is what we will do from now on.
\paragraph{Orbifold}
The superconformal index of the $\mathbb{Z}_k$ orbifold of this LG model is given below. As previously explained, it suffices to  consider the case without superpotential. The index is computed by summing over all flat $\mathbb{Z}_k$ bundles. They are labeled by holonomies $\gamma,\gamma^\prime \in \mathbb{Z}_k$ \cite{Berglund:1994zg,Berglund:1993fj}. The superconformal index of the orbifolded LG model is
\begin{equation}\label{LGorbifold}
{\cal I}_{\bbC^N/\mathbb{Z}_k}(a_i,q,t)\;=\; \frac{1}{k}\sum_{a,b=0}^{k-1} \prod_{i=1}^{N}\Delta(\omega_k^a \,q^{b/k}a_i;q,t),
\end{equation}
where $\omega_k$ is the primitive $k$-th root of unity.

\subsection*{Index of the vector multiplet}
In addition to chiral multiplets, an important ingredient for $(2,2)$ gauge theories is the vector multiplet. The gauge invariant field content of the vector multiplet is given by its field strength multiplet. After dualizing, this multiplet becomes the twisted chiral multiplet $\sigma$. In this paper, we will be considering superconformal gauge theories with a linear twisted superpotential  ${\tilde W}=\zeta\,{\rm tr}\sigma$. The marginality of this interaction determines the r-charge of $\sigma$, $J_{\ttL}=1$ and hence $H_{\ttL}=\frac12$. The single letter index of the vector multiplet $f_V$ can  be obtained by listing the letters of $\sigma$ multiplet with $\delta=0$. We have done that in table \ref{vectorletters}.
\begin{table}
\begin{centering}
\begin{tabular}{|r|r|r|r|r|}
\hline
 & $ H_\ttL $ & $J_\ttL $ & index\tabularnewline
\hline
\hline
$\sigma$ & $\frac{1}{2}$ & $1$ &  $q^{\frac{1}{2}}y$\tabularnewline
\hline
${\bar \sigma}$ & $\frac{1}{2}$ & $-1$ & $q^{\frac{1}{2}}/y$\tabularnewline
\hline
$\chi_{+}$ & $1$ & $0$ & $-q$\tabularnewline
\hline
${\bar \chi}_{+}$ & $1$ & $0$ &  $-q$\tabularnewline
\hline
\hline
$\partial$ & $1$ & $0$ &  $q$\tabularnewline
\hline
\end{tabular}
\par\end{centering}
\caption{\label{vectorletters}The gauge invariant field content of the vector multiplet is encoded in its field strength multiplet. This multiplet is a twisted chiral multiplet with primary $\sigma$.}
\end{table}
Summing the contributions, we get
\be\label{lettervector}
f_V(q,t)=\frac{-2q+t+\frac{q}{t}}{1-q}.
\ee
The vector multiplet index $\II_V(q,t)$ is obtained by its plethystic exponent.  

We can also get the vector multiplet index $\II_V(q,t)$ via the following shortcut. Consider the $U(1)$ gauge theory coupled to a single chiral multiplet. After super-Higgs mechanism, the vector multiplet ``eats" the chiral multiplet to become massive. This results in the empty theory at low energies with superconformal index $1$. As far as the superconformal index is concerned, the Higgsing is implemented by picking the residue at $a=1$. Here $a$ is the fugacity associated with the $U(1)$ gauge symmetry. This prescription is motivated by the similar Higgsing prescription for $4d$ theories as discussed in \cite{Gaiotto:2012xa}.
\be\label{vectorresidue}
1={\rm Res}_{a=1}\Delta(a;q,t){\cal I}_{V}(q,t) \qquad \Rightarrow\qquad {\cal I}_{V}(q,t)=\frac{(q;q)^{2}}{\theta(t;q)}.
\ee
Indeed we see that ${\cal I}_{V}(q,t)=\mathtt{PE}[f_V(q,t)]$.
The index of the non-abelian vector multiplet, $U(k)$ for instance, is obtained by introducing the fugacities  $a_i$ for all the Cartans of $U(k)$. The single letter index of the $U(k)$ vector mutiplet is $f_V(q,t)\chi_{\rm adj}(a_i)$. Taking the plethystic exponent we get,
\be\label{UNvector}
\II_V^{U(k)}(a_i;q,t)=\Big(\frac{(q;q)^{2}}{\theta(t;q)}\Big)^{k}\prod_{i\neq j}\Big((1-\frac{a_i}{a_j}) \Delta(\frac{a_i}{a_j};q,t)\Big)^{-1}.
\ee

\section{Index of abelian gauge Theories}\label{abelian}

In this section we start our investigation of the superconformal index of the $(2,2)$ gauge theories.
The $(2,2)$ supersymmetric gauge theories in $2$ dimensions are constructed out of chiral multiplets $\Phi$ and vector multiplets $V$. For a given field content, the $(2,2)$ theory is labeled by the superpotential $W$, a holomorphic function of chiral superfields. As advocated below \eqref{LGW} we will be considering the case with $W=0$. For each abelian factor in the gauge group we will add the FI term $\zeta \Sigma$ where $\Sigma$ is the corresponding twisted chiral field-strength multiplet.

As detailed in \cite{Witten:1993yc}, when the  $U(1)_A$ R-symmetry is non-anomalous, the beta function for $\zeta$ vanishes and the gauge theory leads to a family of conformal field theories labeled by $\zeta$. For $\text{Im} \zeta \to \infty$, the moduli space of vacua of the gauge theory is given by a Calabi Yau $X$. In this limit, the gauge theory is described as a nonlinear sigma model on the moduli space $X$. On the other hand, as $\text{Im} \zeta \to -\infty$, the moduli space is given by Calabi Yau $\tilde X$ which is generically different from $X$. In this limit the gauge theory has the description in terms of nonlinear sigma model on $\tilde X$. The sigma models on $X$ and $\tilde X$ are  two \emph{phases} of the same theory, they are continuously connected as we go from $\text{Im} \zeta \to \infty$ to $\text{Im} \zeta \to -\infty$.

\subsection*{Gauge theory index}

As emphasized earlier, the superconformal index of the gauge theories can be computed in the free limit. We multiply the index contribution of all the multiplets of the theory to get the index over the entire Fock space. Then we impose the Gauss law by integrating it over the gauge group. This gauge group integral can be reduced to integral over the Cartan torus i.e. to the integral over the gauge fugacities, at the expense of introducing the van der Monde determinant measure $\prod_{i\neq j}(1-a_i/a_j)$. Any gauge group integral is always accompanied by the corresponding vector multiplet index \eqref{UNvector}. The van der Monde determinant cancels with the same factor in \eqref{UNvector} to give us the integral:
\be\label{zintegral}
\II=\oint [dU(k)] \II_V^{U(k)}\ldots = \frac{1}{k!} \Big(\frac{(q;q)^{2}}{\theta(t;q)}\Big)^{k}\oint \prod_i \frac{da_i}{2\pi i a_i} \prod_{i\neq j}\frac{1}{\Delta(a_i/a_j;q,t)}\ldots.
\ee
We effectively get an integral over the Cartan torus with a new measure. The $\ldots$ denotes the contribution of the matter multiplets to the index. This integral is reminiscent of the one computing superconformal index of four dimensional gauge theories \cite{Dolan:2008qi,Gadde:2009kb}.

In this section, we utilize this prescription to compute the superconformal index of the abelian gauge theories. Before we proceed to concrete computation, it will be useful to note the analytic structure of the chiral multiplet index $\Delta(z;q,t)$.
\begin{equation}
\Delta(z;q,t)=\frac{\theta(t;q)}{(q;q)^{2}}\sum_{i\in {\mathbb Z}}\frac{t^{i}}{1-zq^{i}}.
\end{equation}
Note that $\Delta(z;t,q)$ has a pole at $z=q^j$ for all $j\in \mathbb{Z}$.
We will perform the fugacity integrals by evaluating the residues using above expression for $\Delta$.

\subsection*{Residue prescription}

\begin{figure}
\centering
\includegraphics[scale=0.3]{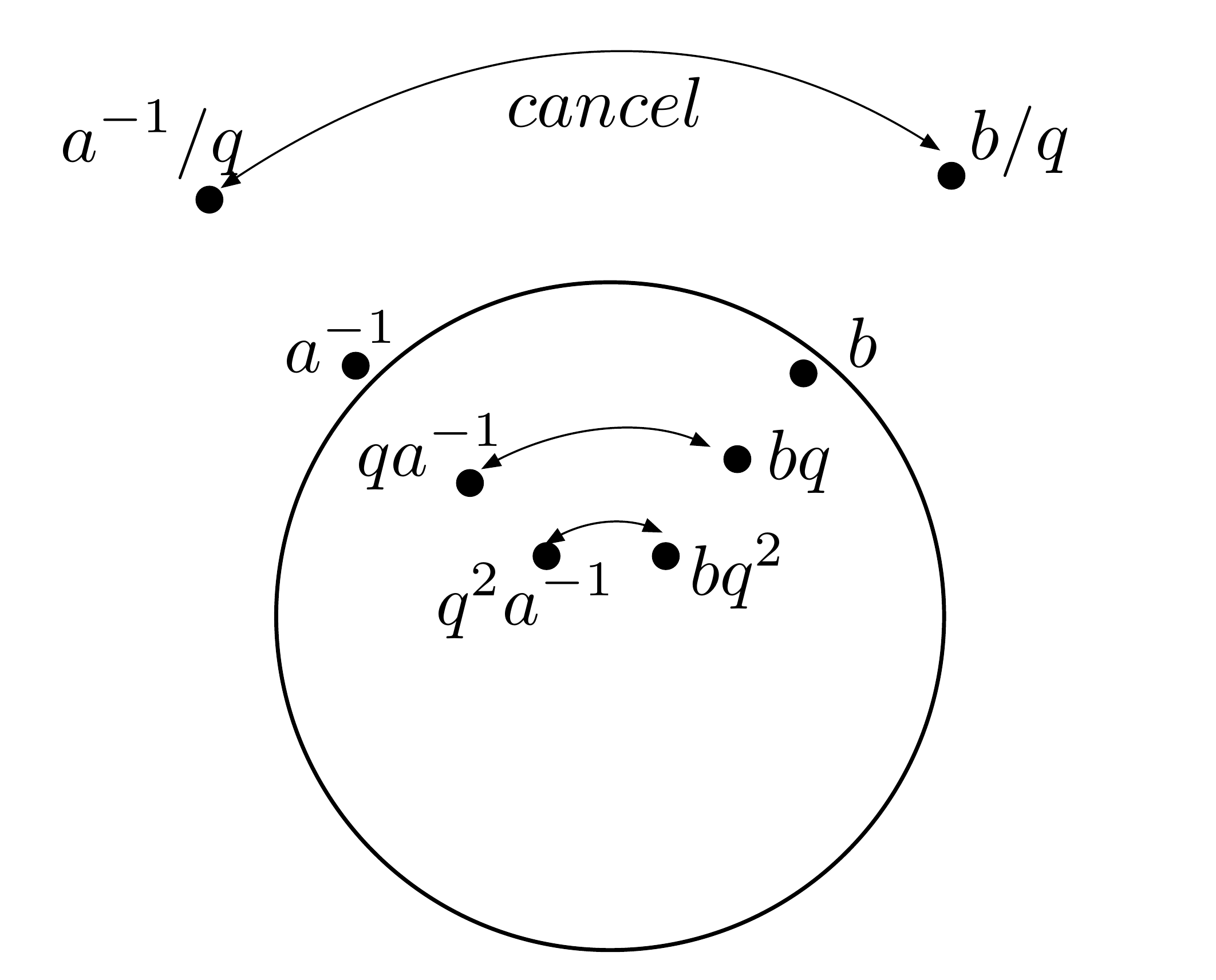}
\caption{The pole structure of the integrand. The residues at poles $z=a^{-1}$ and $z=b$ sum to zero. The arrows indicate canceling contributions to the integral along the unit circle.}
\label{poles}
\end{figure}

Consider a $U(1)$ gauge theory with $N$ chiral mutliplets of charge $Q_i$. The condition for the cancellation of $U(1)_A$ anomaly is $\sum_{i=1}^N Q_i=0$. Using \eqref{zintegral}, the superconformal index of this gauge theory is
\begin{equation}
{\cal I}(a_i;q,t)=\frac{(q,q)^2}{\theta(t,q)}\oint \frac{dz}{2\pi i z} \prod_{i=1}^{N} \Delta(z^{Q_i} a_i;q,t)=:   \oint \frac{dz}{2\pi i z} {\cal J}(z,a_i; q,t).
\end{equation}
where $a_i$ is the fugacity for $U(1)_i$ symmetry rotating the $i$-th chiral mutliplet. Note the transformation of the integrand under $z\to z q$, using \eqref{qshift}
\begin{equation}
{\cal J}(z,a_i;q,t) \to {\cal J}(z q; a_i;q,t)= (1/t)^{\sum_{i=1}^{N} Q_i} {\cal J}(z,a_i;q,t).
\end{equation}
Interestingly $\sum_{i=1}^{N} Q_i=0$ is the condition for the vanishing of $U(1)_A$ anomaly. So for theories flowing to nontrivial SCFTs, the integrand is periodic under $z\to zq$. This means that the integrand is naturally defined on the torus rather than the complex plane. Let us look at the poles of the integrand more closely. As an example, take $N=2$ and $Q_i=\pm1$. Due to the periodicity, the residue at $z= a_1^{-1}$ and at its image $z= a_1^{-1} q^{j}$ are equal i.e. $\text{Res}_{z=a_1^{-1}}=\text{Res}_{z=a_1^{-1}q^j}$. Moreover, as the residue from poles in the fundamental region of the torus should sum to zero,
\begin{equation}\label{magic}
\text{Res}_{z=a_1^{-1}}=-\text{Res}_{z=a_2}.
\end{equation}
Because $|a_i|<1$, the pole at $z=a_2$ is just inside the contour and the pole at $z=a_1^{-1}$ is just outside. The residues at $z=a_1^{-1} q^j$ and at $z=a_2 q^j$ cancel pairwise for each $j$  inside the contour, see figure \ref{poles}.
 We only pick up ${\cal I}=\text{Res}_{z=a_2}$. We can also choose to close the contour to $\infty$ which will pick up $-\text{Res}_{z=a_1^{-1}}$. Both these residues are equal, thanks to \eqref{magic}. In a somewhat ad hoc fashion, we ignore the contribution coming from the singularity at the origin. We justify this prescription by checking it in simple abelian examples. 
 
As we will see later, for non-abelian gauge theories, the index is computed by integrating over the Cartan subgroup of the gauge group. As in the abelian case, the $U(1)_A$ anomaly cancellation condition implies that the integrand is elliptic in all the fugacities. The poles in the fundamental domain of a given fugacity are contributed by the chiral multiplets that are charged positively as well as negatively under the corresponding Cartan. The contour prescription then instructs one to perform the sum of the residues at the positively charged poles only (or equivalently, minus the sum of residue at the negatively charged poles). This procedure is repeated for all the Cartan generators. 
\subsection{Flop transition}

As the first concrete example, consider $U(1)$ gauge theory with $N$ chiral multiplets of charge $+1$ and $N$ chiral multiplet of charge $-1$. The superconformal index is
\begin{equation}
{\cal I}({\bf a},{\bf b};q,t)  =  \frac{(q;q)^{2}}{\theta(t;q)}\oint\frac{dz}{2\pi iz}\prod_{i=1}^{N}\Delta(za_{i};q,t)\Delta(b_{i}/z;q,t)
\end{equation}
The symbols ${\bf a}$ and ${\bf b}$ stand for the collection of variables $\{a_i\}$ and $\{b_i\}$ respectively. The contour prescription gives two different presentations of the index:
\be
{\cal I}_+({\bf a},{\bf b};q,t)& = &\sum_{j=1}^{N}\Delta(a_{j}b_{j};q,t) \prod_{i\neq j}\Delta(a_{i}b_{j};q,t)\Delta(\frac{b_{i}}{b_{j}};q,t)\\
{\cal I}_-({\bf a},{\bf b};q,t)& = &\sum_{j=1}^{N}\Delta(a_{j}b_{j};q,t)\prod_{i\neq j}\Delta(\frac{a_{i}}{a_{j}};q,t)\Delta(b_{i}a_{j};q,t).
\ee
In the first case, we have collected the residues at the poles contributed by negatively charged chiral multiplets,  $z=b_i$ and in the second case, the residues are summed over the poles at $z=a_i^{-1}$ which are contributed by the positively charged chiral multiplets. Interestingly, the poles of this integral are in one to one correspondence with the $N$ vacua of the theory.
This is indeed expected because the residue at each pole contributes $1$ in the limit all fugacities go to zero. In this limit, the superconformal index reduces to the Witten index which counts vacua of the theory (with sign). At low energies, this gauge theory exhibits two phases. In the limit $\text{Im} \zeta \to \infty$, the physics is described by the nonlinear sigma model on $\oplus_{i=1}^N O(-1)_{B} \to \mathbb{CP}_{A}^{N-1}$. On the other hand, in the limit $\text{Im} \zeta\to-\infty$, we get a nonlinear sigma model on $\oplus O(-1)_{A} \to \mathbb{CP}_{B}^{N-1}$. This transition has been called the \emph{flop} transition in the literature. Essentially the base and the fiber of the Calabi-Yau swap their roles, i.e. $A\leftrightarrow B$. Because the indices $\II_+$ and $\II_-$ are also related by ${\bf a} \leftrightarrow {\bf b}$, we interpret $\II_+$ and $\II_-$ to be the superconformal indices of the two phases. They are manifestly equal.

Phase transition under the sign change of $\text{Im}\zeta$ is a generic feature of $(2,2)$ abelian gauge theories. We will see that the
lesson learned in this example holds in general: the index of the phase obtained in the $\text{Im} \zeta \to \infty$ limit is computed by summing  the residues of the poles coming from positively charged chiral multiplets while the index of the other phase is obtained by summing the residues coming from negatively charged chiral multiplets. They are equal by construction. In this example, the two phases had very similar sigma model descriptions. That is not the case in general. Next example illustrates this point to a greater effect.

\subsection{CY-LG correspondence}

Consider a $U(1)$ gauge theory with $N$ chiral field of charge $-1$ and $1$ chiral field of charge $+N$. The $U(1)_A$ anomaly vanishes as the gauge charges  sum to zero. The theory flows to a nontrivial fixed point labeled by $\zeta$. For $\text{Im} \zeta \to \infty$, we get a sigma model on $O(-N) \to \mathbb{CP}^{N-1}$ and for $\text{Im} \zeta \to -\infty$, we get a $\bbC^N/{\mathbb Z}_N$ Landau-Ginzburg model. This is the classic case of the so called \emph{CY-LG correspondence} discovered in \cite{Witten:1993yc}. The superconformal index of the UV gauge theory is computed by the integral,
\begin{equation}
{\cal I}({\bf a},{b};q,t)   =  \frac{(q;q)^{2}}{\theta(t;q)}\oint\frac{dz}{2\pi iz}\Delta(b z^N;q,t)\prod_{i=1}^{N}\Delta(a_{i}/z;q,t).
\end{equation}
Evaluating it in two ways as before, we get
\be
{\cal I}_+({\bf a},{b};q,t) &  =  & \sum_{j=1}^{N}\Delta(b a_j^N;q,t)\prod_{i\neq j}\Delta(\frac{a_{i}}{a{j}};q,t)\\
{\cal I}_-({\bf a},{b};q,t) & = & \frac{1}{N} \sum_{j,k=0}^{N-1}  \prod_{i=1}^{N}\Delta(\omega_N^j q^{\frac{k}{N}} b^{\frac{1}{N}}a_{i};q,t).
\ee
From \eqref{LGorbifold} we see that $\II_-$ is precisely the index of $\bbC^N/{\mathbb Z}_N$ Landau-Ginzburg orbifold, one phase of the theory. Hence we claim that $\II_+$ is the index of the other phase i.e. of the nonlinear sigma model on $O(-N) \to \mathbb{CP}^{N-1}$. Again, $\II_+=\II_-$ by construction.

It is straightforward to generalize the discussion in this section to theories with multiple $U(1)$ gauge groups. Depending on the values of the FI parameters, the gauge theory engineers a sigma model on a toric Calabi-Yau. All phases of this theory are related by what is known as the toric-duality. Our treatment of the superconformal index automatically implies that the superconformal index of all the toric dual Calabi-Yau manifolds is equal.

\section{Non-abelian gauge theories}\label{nonabelian}

Now we turn our attention to non-abelian superconformal gauge theories. The focal point of this section will be a $U(k)$ gauge theory with $N$ fundamental chiral multiplets and other chiral multiplets required to cancel $U(1)_A$ anomaly. Depending on whether the matter content includes an adjoint chiral field or not, these theories can be divided into two classes. The motivation for this division  comes from the fact that vacuum classification for these two types of theories is vastly different. For theories without adjoint chiral multiplet, the vacua correspond to $k$ choices out of $N$ while for the theories with adjoint chiral multiplets, the vacua are classified by length $N$ partitions of $k$. The carefully engineered theories of the former class exhibit interesting Seiberg-type duality under the exchange of gauge group $U(k)\leftrightarrow U(N-k)$. This is consistent as the number of vacua $N\choose k$ is symmetric under $k\leftrightarrow N-k$. Some of these dualities were first discovered by Hori and Tong \cite{Hori:2006dk}. We will first analyze theories without adjoint chiral multiplet and then consider the theories with adjoint chiral multiplet. The later will play an important role in section \ref{surfaceindex}.

\subsection{Without adjoint matter and Hori-Tong duality}\label{nonabelian-without}

The moduli space of a $U(k)$ gauge theory with $N$ fundamental chiral multiplets is the Grassmannian $\mbox{Gr}(k,N)$. Inspired by the the four dimensional Seiberg duality and the equivalence $\mbox{Gr}(k,N)\simeq \mbox{Gr}(N-k,N)$, this theory is conjectured to be dual to $U(N-k)$ gauge theory with $N$ fundamental chiral fields. The $U(1)_A$ symmetry is anomalous in these theories. We are interested in studying the superconformal index hence we would like to cancel the anomaly to get an SCFT. Let us look at the following ways of doing so.
\begin{enumerate}
\item Add $\ell$ chiral fields, $i$-th field transforming in det$^{-q_i}$ representation s.t. $\sum_{i=1}^{\ell}q_i=N$.\label{det-option}
\item Add $N$ chiral fields transforming in the anti-fundamental representation.\label{anti-option}
\item Use $SU(k)$ gauge group instead of $U(k)$.\label{S-option}
\end{enumerate}
Adopting these three options gives us three different versions of the Seiberg-type duality for superconformal theories. We will analyze all of them one by one.

\subsection*{Version 1}
In this subsection, we consider the option \ref{det-option} and demonstrate that the superconformal index is equal on both sides of the duality. We will take $\ell=1$ for definiteness, generalization to arbitrary $\ell$ is straightforward. Consider the $U(k)$ gauge theory with $N$ fundamental chiral fields and $1$ chiral field in transforming in det$^{-N}$ representation. Its index is computed by the integral:
\be
\II_{k}^{(1)}(a,{\bf b};q,t)=\Big(\frac{(q;q)^2}{\theta(t,q)}\Big)^{k} \frac{1}{k!}\oint\prod_{\alpha=1}^{k}\frac{dz_\alpha}{2\pi iz_\alpha}
\frac{\Delta(a(\prod  z)^{-N};q,t)\prod_{\alpha=1}^{k}\prod_{j=1}^{N}\Delta(b_{j}/z_\alpha;q,t)}{\prod_{\alpha\neq \beta}\Delta(z_{\alpha}/z_{\beta};q,t)}.
\ee
We have introduced fugacities $b_i$ for the $SU(N)$ flavor symmetry of $N$ fundamental chiral multiplets. They obey $\prod_i b_i=1$. The fugacity $a$ is for the $U(1)$ symmetry rotating the single remaining chiral multiplet. The denominator is the contribution of the vector multiplet. We pick the residues at the poles coming from negatively charged chirals,  $z_\alpha=b_{i_\alpha}$. This choice of the pole is indicated by the subset $\{i_\alpha:\alpha=1,\ldots, k\}\subset\{i:i=1,\ldots, N\}$. As is the case with abelian gauge theories, the poles of the nonabelian index integral are also in one to one correspondence with the vacua of the theory. The fact that there are $N\choose k$ number of poles is promising because this number is symmetric under $k\leftrightarrow N-k$. Summing over all the residues,
\be
\II_k^{(1)}(a,{\bf b};q,t)&=&\sum_{\{i_\alpha\}} \frac{\Delta(a(\prod_\alpha  b_{i_\alpha})^{-N};q,t)\prod_{\alpha=1}^{k}\prod_{j\neq i_\alpha}\Delta(b_{j}/b_{i_\alpha};q,t)}{\prod_{\alpha\neq \beta}\Delta(b_{i_\alpha}/b_{i_\beta};q,t)}\notag\\
&=& \sum_{\{i_\alpha\}} \Delta(a(\prod_{s\in \{i_\alpha\}}  b_s)^{-N};q,t) \prod_{s\in \{i_\alpha\}} \prod_{r\in \bar{\{i_{\alpha}\}}}\Delta(b_r/b_s;q,t).
\ee
The subset $\bar{\{i_{\alpha}\}}$ is the complement of the subset $\{i_\alpha\}$. It is easy to see that the second line is completely symmetric under the exchange $(\{i_\alpha\}\leftrightarrow \bar{\{i_{\alpha}\}}, b_{i} \leftrightarrow {\tilde b_i}:=b_i^{-1})$. This leads to the identity,
\be\label{HT1}
\II_k^{(1)}(a,{\bf b};q,t)=\II_{N-k}^{(1)}(a,{\tilde{\bf b}};q,t).
\ee
The equation \eqref{HT1} serves as a powerful check of the duality version \ref{det-option}.

\subsection*{Version 2}

In this section we analyze the variation of the Seiberg-type duality for the superconformal field theory resulting from adding $N$ anti-fundamental chiral  fields to the $U(k)$ gauge theory with $N$ fundamental chiral fields. This is the electric side of the duality. In this case we will see that the magnetic dual theory doesn't just involve changing of gauge group to $U(N-k)$ but also the addition of meson fields coupled to the matter field via cubic superpotential, reminiscent of the Seiberg duality in four dimensions. The index of the electric side is computed by the integral,
\be
\II_{k}^{(2)}({\bf a},{\bf b},c;q,t)=\Big(\frac{(q;q)^2}{\theta(t,q)}\Big)^{k} \frac{1}{k!}\oint\prod_{\alpha=1}^{k}\frac{dz_\alpha}{2\pi iz_\alpha}
\frac{\prod_{\alpha=1}^{k}\prod_{j=1}^{N}\Delta(cz_{\alpha}a_{j};q,t)\Delta(cb_{j}/z_\alpha;q,t)}{\prod_{\alpha\neq \beta}\Delta(z_{\alpha}/z_{\beta};q,t)}.
\ee
The variables ${\bf a},{\bf b}$ are the fugacities for $SU(N)_A\times SU(N)_B$ flavor symmetry while $c$ is the fugacity for the (relative) $U(1)$ symmetry.
Again the poles of this integral are at $z_\alpha=b_{i_\alpha}$. The poles are classified by the subset $\{i_\alpha\}$ as before.
\be
\II_k^{(2)}({\bf a},{\bf b};q,t)&=&\sum_{\{i_\alpha\}} \frac{ \prod_{\alpha=1}^{k}\prod_{j}\Delta(c^2 a_j b_{i_\alpha};q,t)}{\prod_{\alpha\neq \beta}\Delta(b_{i_\alpha}/b_{i_\beta};q,t)}\prod_{\alpha=1}^{k}\prod_{j\neq i_\alpha}\Delta(b_{j}/b_{i_\alpha};q,t)\notag\\
&=& \sum_{\{i_\alpha\}} \prod_{s\in \{i_\alpha\}}  \prod_j \Delta(c^2 a_j b_s;q,t)\prod_{s\in \{i_\alpha\}}  \prod_{r\in \bar{\{i_{\alpha}\}}}\Delta(b_r/b_s;q,t).
\ee
The first factor in the summation can be written as
\be
\prod_{s\in \{i_\alpha\}}  \Delta(c^2 a_j b_s;q,t)
=\frac{\prod_{i}  \Delta(c^2 a_j b_i;q,t)}{\prod_{r\in \bar {\{i_\alpha\}}}   \Delta(c^2 a_j b_r;q,t)}
= \prod_{i} \Delta(c^2 a_j b_i;q,t) \prod_{r\in \bar {\{i_\alpha\}}} \Delta(\frac{qt^{-1}}{c^2}\frac{1}{a_j}\frac{1}{ b_r};q,t).\nonumber
\ee
In the second equation we have used the property \eqref{Deltainverse}. This allows us to write the equality,
\be\label{HT2}
\II_k^{(2)}({\bf a},{\bf b},c;q,t)= \Big(\prod_{i,j}\Delta(\frac{qt^{-1}}{{\tilde c}^2}\frac{1}{{\tilde a}_i {\tilde b}_j};q,t)\Big) \,\,\II_{N-k}^{(2)}({\tilde{\bf a}},{\tilde{\bf b}},{\tilde c};q,t)
\ee
where ${\tilde a}_i:=1/a_i, {\tilde b}_i:=1/b_i$ and ${\tilde c}=\sqrt{qt^{-1}}/c$. The right hand side is the index of $U(N-k)$ gauge theory with $N$ fundamental chiral fields $q_i$ and $N$ anti-fundamental chiral fields ${\tilde q}_j$ coupled to $N^2$ gauge singlet meson fields $M_{ij}$ through the superpotential $W=q_i M_{ij} {\tilde q}_j$. The pre-factor on the RHS is the index contribution of the meson fields. The power of $qt^{-1}$ appearing in the argument of the $\Delta$ function contribution of the meson is due to the fact that it has r-charge $1$. The equation \eqref{HT2} offers a powerful check of the duality version \ref{anti-option}. This theory will play an important role in section \ref{dualitycheck} where we compute the index of half-BPS surface operators in $\CN=2$ superconformal gauge theories.

\subsection*{Version 3}

Next we consider the option \ref{S-option}. Although an explicit computation of the index could be done in this case as well, we will argue the index equality by showing that index of both sides satisfies the same $q$-difference equation. In doing so we will introduce an important physical concept that will be studied in section \ref{physical}. The index of $SU(k)$ theory with $N$ fundamental chiral fields is computed by the integral:
\be
\II_{k}^{(3)}({\bf b};q,t)=\Big(\frac{(q;q)^2}{\theta(t,q)}\Big)^{k-1} \frac{1}{k!}\oint _{\prod z=1}\prod_{\alpha=1}^{k-1}\frac{dz_\alpha}{2\pi iz_\alpha}
\frac{\prod_{\alpha=1}^{k}\prod_{j=1}^{N}\Delta(b_{j}/z_\alpha;q,t)}{\prod_{\alpha\neq \beta}\Delta(z_{\alpha}/z_{\beta};q,t)}.
\ee
Note that the integral over $z_i,\,i=1,\ldots, k$ is constrained with $\prod z=1$ because $z_i$ are $SU(k)$ fugacities. 
The fugacities $b_i$ couple to the Cartan of the $U(N)$ flavor symmetry. Let us introduce the operators $p_{b_i}$ and $p_{t}$ that obey the $q$-commutation $p_x x=q\, x p_x$ for $x=b_i,t$. Using \eqref{Deltashift} we see that the integral $\II_k^{(3)}({\bf b};q,t)$ obeys simple difference equations,
\be\label{diffeq}
p_{b_i}-(1/t)^k=0,\qquad\qquad p_t-(-1/t)^{k(N-k)+1}\prod_i b_i^{-k}=0.
\ee
Now consider the integral $\II_{N-k}^{(3)}({\tilde {\bf b}};q,t)$, ${\tilde b}_i:=(\prod b)^{1/N-k}/b_i$. It easy to see that this integral satisfies the same $p_t$ difference equation. It is somewhat subtle to compute the action of $p_{b_i}$ on this integral. In what follows, we will compute this action and show that $\II_{N-k}^{(3)}({\tilde {\bf b}};q,t)$ also satisfies the $p_{b_i}$ difference equation \eqref{diffeq}. Note that
\begin{eqnarray}
p_{b_i}\tilde{b}_{j} & = & q^{\frac{1}{N-k}-1}\tilde{b}_{j}p_{b_i}\qquad\mbox{for}\quad i=j\nonumber \\
 & = & q^{\frac{1}{N-k}}\tilde{b}_{j}p_{b_i}\qquad\quad\mbox{for}\quad i\neq j.
\end{eqnarray}
The fractional power shifts in the flavor fugacities make it harder to deal with the $p_{b_i}$ action on the integral $\II_{N-k}^{(3)}({\tilde {\bf b}};q,t)$ directly. We take an indirect approach. After the action of $p_{b_i}$ on $\II_{N-k}^{(3)}({\tilde {\bf b}};q,t)$, the shifts $q^{1/{N-k}}$ in the flavor fugacity $\tilde{b}_{j}$, for all $j$, are absorbed into the magnetic gauge fugacity $\tilde z_{\alpha}\to q^{1/{N-k}}\tilde z_{\alpha}$. As a result, instead of obeying $\prod_{\alpha=1}^{N-k}\tilde z_\alpha=1$ the contour obeys $\prod_{\alpha=1}^{N-k}\tilde z_\alpha=q$. The remaining shift of $q^{-1}$ in $\tilde{b}_{i}$, gives rise to the overall factor $(1/t)^{k-N}$. We get,
\begin{equation}
p_{b_i}\II_{N-k}^{(3)}({\tilde {\bf b}};q,t)=(1/t)^{k-N} \II_{N-k}^{(3)}({\tilde {\bf b}};q,t)|_{\prod z=q}.
\end{equation}
The new $q$-contour can be changed back to the unit contour by rescaling one of the $\tilde z_{\alpha}$ variables as $\tilde z_{\alpha}\to \tilde z_{\alpha}q$ and absorbing this shift back in the flavor fugacities $\tilde b_i \to q\tilde b_i$. This shift gives rise to the additional factor $(1/t)^N$, showing that $\II_{N-k}^{(3)}({\tilde {\bf b}};q,t)$ obeys the same $p_{b_i}$ difference equation as \eqref{diffeq}.

\subsection{With adjoint matter}\label{nonabelian-with}
As a final example, consider a $U(k)$ gauge theory with $N$ fundamental, $N$ anti-fundamental chiral multiplets and one adjoint chiral multiplet $\varphi$. The flavor symmetry of this theory is $SU(N)_A\times SU(N)_B \times U(1)_c \times U(1)_d$. The symmetries $SU(N)_A$ and $SU(N)_B$ act on fundamental and anti-fundamental chiral multiplet respectively. The $U(1)_c$ symmetry is the overall symmetry rotating the fundamental and anti-fundamental fields in the opposite fashion while the $U(1)_d$ acts only on the adjoint chiral fields. This theory will play a crucial role in section \ref{vortex} where we consider the surface operators coming from vortex strings. The index is give by the integral,
\be
&&\II_{k}^{(4)}({\bf a},{\bf b},c,d;q,t)=\notag\\
&&\Big(\frac{(q;q)^2}{\theta(t,q)}\Big)^{k} \frac{1}{k!}\oint\prod_{\alpha=1}^{k}\frac{dz_\alpha}{2\pi iz_\alpha}
\frac{\prod_{\alpha, \beta}\Delta(d\frac{z_\alpha}{z_\beta};q,t)}{\prod_{\alpha\neq \beta}\Delta(\frac{z_\alpha}{z_\beta};q,t)}
\prod_{\alpha=1}^{k}\prod_{j=1}^{N}\Delta(c z_{\alpha}a_{j};q,t)\Delta(c\frac{b_j}{z_\alpha};q,t).
\ee
We have used the fugacities ${\bf a},{\bf b},c$ and $d$ for $SU(N)_A,SU(N)_B,U(1)_c$ and $U(1)_d$ respectively. The pole structure  of this integrand is qualitatively different from the integrands for theories without adjoint chiral multiplet studied in previous subsection. Here they are classified by length $N$ partitions of $k$ i.e. $\{n_i\}$ s.t. $\sum_{i=1}^N n_i= k$.
\be
z_{i,k_{i}}=c b_{i} d^{k_{i}},\qquad k_{i}=0,\ldots,n_{i}-1.
\ee
Each pole corresponds to a Higgs vacuum of the theory. Evaluating the residue at these poles we have,
\begin{eqnarray}\label{withadj}
\II_{k}^{(4)}({\bf a},{\bf b},c,d;q,t)& = & \sum_{\{n_i\}}\prod_{i,j}^{N}\prod_{n=0}^{n_{i}-1}\prod_{l=0}^{n_{j}-1}\frac{\Delta(d^{1+l-n}b_{j}/b_{i})}{\Delta(d^{l-n}b_{j}/b_{i})}\prod_{i,j}^{N}\prod_{n=0}^{n_{i}-1}\Delta(c^{2}a_{j}b_{i}d^{n})\Delta(d^{-n}b_{j}/b_{i})\nonumber\\
 & = &\sum_{\{n_i\}} \prod_{i,j}^{N}\prod_{n=0}^{n_{i}-1}\Delta(c^{2}a_{j}b_{i}d^{n})\Delta(d^{n_{j}-n}b_{j}/b_{i}).
\end{eqnarray}
This provides us with the explicit expression for the superconformal index of this gauge theory. We will borrow this result in section \ref{surfaceindex}. The eq. \eqref{withadj} is not symmetric under the exchange $k\leftrightarrow N-k$ hence we do not expect a  Seiberg-type type duality in this case. On the other hand if a superpotential  $W=\mbox{tr} \,\varphi^{\ell+1}$ is added for the adjoint chiral multiplet,  
inspired from Kutasov-Schwimmer duality in four dimensions \cite{Kutasov:1995np}, we expect this theory to be dual to $U(\ell N-k)$ gauge theory with the same matter content and $\ell$ mesons.  Note that for $\ell=2$ one can integrate out the adjoint chiral multiplet then this duality reduces to Seiberg type duality discussed in \ref{nonabelian-without} version 2. It would be interesting to pursue this direction further.

\subsection{$q$-difference equations}\label{physical}

The difference equations \eqref{diffeq} that we used for identifying indices of dual 2d gauge theories
have a similar form to the $q$-difference operators which annihilate various partition functions of
3d gauge theories with the same amount of supersymmetry (namely, four real supercharges) \cite{Dimofte:2011jd,Dimofte:2011py}.
The latter can be understood either as Ward identities for line operators or, alternatively,
as quantum operators produced via quantization of moduli spaces of SUSY parameters after
the theory is compactified on a circle, so that the theory is also effectively two-dimensional \cite{Dimofte:2010tz}.

Our goal here is to understand the origin of the analogous quantum operators that annihilate the flavored elliptic genus.
In particular, as we explain below, operator equations like \eqref{diffeq} is a general feature of 2d $\CN=(2,2)$ theories as well.
However, their origin and interpretation are qualitatively different from what one finds in (circle reductions of) 3d theories.

In general, starting with a Cartan torus $\mathbb{T}$ of the flavor symmetry group,
we introduce the so-called ``quantum torus,'' {\it i.e.} the set of operators $x_i$ and $p_i$ that obey
\beq
p_i x_i \; = \; q x_i p_i
\label{pqcomm}
\eeq
Moreover, we will think of these as operators acting on functions of $x_i$, so that each $p_i$ is a ``shift operator.''
Note, the 2d index $\II ({\bf a};q,t)$ is an example of such a function, where the set $\{ x_i \}$
comprises the flavor fugacities $\{ a_i \}$ as well as the (modified) Jacobi variable $t$.

In order to understand the meaning of the $q$-difference operators $\hat A_i ({\bf x}, {\bf p}, q)$
annihilating the 2d index, it is convenient to consider the classical limit $q = e^{\hbar} \to 1$,
in which quantum operators are replaced by classical equations defining an algebraic variety, {\it cf.} \cite{Gukov:2003na}:
\beq
\hat A_i ({\bf x}, {\bf p}, q) \, \II \; = \; 0
\qquad
\stackrel{q \to 1}{\rightsquigarrow}
\qquad A_i ({\bf x}, {\bf p}) \; = \; 0
\label{AAclq}
\eeq
For example, the classical equations associated with \eqref{diffeq} have the form
\beq
A_i \; = \; p_{b_i} - t^{-k}
\eeq
and similarly for the pair of conjugate variables $(t,p_t)$, which we ignore for now to avoid clutter.

These classical equations, then, control the asymptotic behavior of the 2d index in the limit $q = e^{\hbar} \to 1$,
which turns out to be very simple:
\beq
\II ({\bf a};q,t) \; \underset{q \to 1}{\simeq} \; \exp \left( \frac{1}{\hbar} \CW ({\bf a}, t) + \ldots \right)
\label{indlim}
\eeq
Namely, for many 2d $\CN=(2,2)$ theories with a Lagrangian description, the function $\CW ({\bf a},t)$
--- which plays the role analogous to the twisted superpotential in 3d theories compactified on a circle ---
turns out to be a quadratic function of $\log a_i$ and $\log t$.
Moreover, the dependence on $\log a_i$ is actually {\it linear},
\beq
\CW \; = \; - \log t \sum_i n_i \log a_i + \ldots
\label{Wlinq}
\eeq
where the ``anomaly coefficient'' $n_i$ counts (with signs) all charges of matter fields
in the theory under the $i$-th flavor symmetry.
Before we explain the origin of \eqref{Wlinq}, let us see what the consequences of this simple structure are.
Substituting it into \eqref{indlim} and thinking about the classical limit of the operators \eqref{pqcomm}
acting on the 2d index, one quickly finds that the algebraic variety defined by the equations $A_i= 0$
is simply a graph of gradient of the ``potential'' $\CW$:
\beq
A_i ({\bf x}, {\bf p}) \; = \; 0
\qquad
\Leftrightarrow
\qquad
p_i \; = \; \exp \left(\frac{\partial \CW}{\partial \log x_i} \right)
\label{AviaW}
\eeq
In particular, for $x_i = a_i$ this gives
\beq
A_i \; = \; p_i - t^{-n_i}
\eeq
where we used \eqref{Wlinq}.

Now, let us justify \eqref{Wlinq} by analyzing the $q \to 1$ limit of the 2d index.
Since basic building blocks of general 2d $\CN=(2,2)$ theories are gauge vector multiplets and
chiral matter multiplets, it is essential to understand the asymptotic behavior of the chiral multiplet index
$\Delta(a;q,t)$ in this limit.
Since $\Delta(a;q,t)$ is built from $\theta$-functions which, in turn, can be written
as ratios of two $q$-Pochhammer symbols, we will need the following facts in this analysis:
\beq
(x;q)_n \; = \; \prod_{i=0}^{n-1} (1 - x q^i) \; \underset{q \to 1}{\simeq} \; e^{\frac{1}{\hbar} \left( \text{Li}_2 (x) - \text{Li}_2 (x q^n) \right)}
\eeq
and
$$
\text{Li}_2 (x) \; = \; - \text{Li}_2 (x^{-1}) - \frac{1}{2} \left[ \log (-x) \right]^2 - \frac{\pi^2}{6}
\; = \; - \text{Li}_2 (x^{-1}) - \frac{1}{2} (\log x)^2 - i \pi \log x + \frac{\pi^2}{3}
$$
Using these relations, it is easy to see that in the limit $q = e^{\hbar} \to 1$,
\beq
\theta (x;q) \simeq e^{\frac{1}{\hbar} \left( \text{Li}_2 (x) + \text{Li}_2 (x^{-1}) \right) + \ldots}
\eeq
and, finally,
\begin{align}
\Delta (a;q,t)
& = \frac{\theta (at;q)}{\theta (a^{-1};q)} \notag \\
& \simeq e^{\frac{1}{\hbar} \left( \text{Li}_2 (at) + \text{Li}_2 (a^{-1}t^{-1}) - \text{Li}_2 (a) - \text{Li}_2 (a^{-1}) \right) + \ldots} \label{Dchasympt} \\
& = e^{\frac{1}{\hbar} \left( - \log (-a) \log t - \frac{1}{2} (\log t)^2 \right) + \ldots} \notag
\end{align}
In particular, we conclude that the asymptotic behavior of the 2d index indeed
has the proposed form \eqref{indlim}, with linear dependence of $\CW$ on $\log a$.
Specifically, each chiral multiplet with charge $+1$ under a global $U(1)$ symmetry group
contributes $\Delta \CW = - \log t \cdot \log z$, where $a$ is the fugacity for that symmetry.
Using similar arguments, one finds that in a theory with several chiral multiplets with
various charges the behavior of the 2d index in the limit $q \to 1$ has a simple form \eqref{indlim}--\eqref{Wlinq}.

Gauging some of the flavor symmetries leads to 2d gauge theories, whose index is given by
the integral \eqref{zintegral} over the Cartan torus.
Potentially, this can alter the simple form \eqref{Wlinq} of the 2d index in the limit $q \to 1$
and, therefore, lead to non-linear classical / quantum relations \eqref{AAclq}.
However, when the saddle point analysis gives a reliable approximation to the integral \eqref{zintegral}
the conclusion \eqref{Wlinq} remains unchanged because extremization of the linear function $\CW$
with respect to the variable $\log a$ associated with a symmetry that is being gauged still
results in a linear function of the remaining variables.
Note, in the case of non-abelian 2d gauge theories, the integration measure $\prod_{i \ne j} \Delta(a_i/a_j; q,t)^{-1}$
does not present a problem since it is still a product of factors \eqref{Dchasympt}, all of which yield a linear potential $\CW$.
In the examples of gauge theory index computations presented in section \ref{abelian} and \ref{nonabelian}, this can also be seen by noticing that the final integral is given by sum of products of $\Delta$ function.

Returning to our example of $SU(k)$ gauge theory with $N$ fundamental chiral multiplets
discussed in the previous section, we conclude that the 2d index $\II_{k}^{(3)}({\bf b};q,t)$
has the asymptotic form \eqref{indlim} with
\beq
\CW \; = \; - k \log t \sum_i \log b_i + \ldots
\eeq
Then, substituting this into \eqref{AviaW} we find a system of linear constraints with
\be
A_i \; = \; p_{b_i} - t^{-k}
\ee
which is indeed the ``classical'' ($q \to 1$) limit of the difference equations \eqref{diffeq}.

Besides the limit $q \to 1$ discussed in detail here, there are various other limits of the superconformal index
that would be interesting to explore further. For example, one can consider the limit $q \to 0$ which from the viewpoint
of the above discussion would correspond to the ``extreme quantum limit'' $|\hbar| \to \infty$.
In this limit, the $q$-series expansion of the index truncates to a finite polynomial, namely the equivariant
version of the Hirzebruch genus. It captures, for superconformal theories based on sigma-models, the basic
information about  equivariant K-theory of the target manifold.

\section{Coupling to 4d gauge theory}

One prominent feature of all 2d theories considered above is that they all have global (flavor) symmetries.
In fact, it is these symmetries which were  the center of our attention and which allowed us
to introduce the ``flavored'' version of the elliptic genus.
There is a lot more one can do with all such theories, including gauging their flavor symmetries
with --- not necessarily two-dimensional! --- gauge fields; paraphrasing Daniel Defoe, {\it If the shoe fits, wear it.}

In particular, gauging flavor symmetries of a 2d theory $\TS$ with 4d gauge fields yields a two-dimensional
defect, the so-called surface operator, in the four-dimensional gauge theory.

\subsection{Levi types}\label{N4}

We wish to describe a fairly large class of half-BPS surface operators in supersymmetric gauge theories
(with $\CN=4$ as well as $\CN=2$ supersymmetry) that preserve part of the gauge symmetry group $\mathbb{L} \subseteq G$
along the support $S$ of the surface operator.
We shall refer to such surface operators as surface operators of Levi type $\mathbb{L}$ and mostly focus
on the case $G = U(N)$ or $G = SU(N)$, for which different Levi types are classified by partitions of $N$,
\beq
N = \lambda_1 + \lambda_2 + \ldots + \lambda_s
\eeq
For example, the partition $\lambda = [3,1]$ corresponds to the Levi subgroup $\mathbb{L} \subset SU(4)$
that consists of matrices of the form
\beq
\begin{pmatrix}
* & \quad* & \quad0 & \quad0 \\
* & \quad* & \quad0 & \quad0 \\
0 & \quad0 & \quad* & \quad0 \\
0 & \quad0 & \quad0 & \quad*
\end{pmatrix}
\label{Leviex}
\eeq
More generally, the Levi subgroup $\mathbb{L} \subseteq SU(N)$ associated to a partition $\lambda = [\lambda_1, \ldots, \lambda_s]$
has the form
\beq
\mathbb{L} \; = \; S \left( U(k_1) \times \ldots \times U(k_n) \right)
\label{Levilambda}
\eeq
where $k_i$ are the parts, not necessarily in order, of the conjugate ({\it i.e.} transposed) partition $\lambda^t$:
\beq
\lambda^t \; = \; \text{ord} (k_1, \ldots, k_n)
\label{laprel}
\eeq
{}From the Ferrers diagram (a.k.a. Young diagram) it is easy to see that
the $i$-th part of $\lambda^t$ is equal to the number of parts $\ge i$ in $\lambda$,
or equivalently the largest $j$ such that $\lambda_j \ge i$:
\beq
\lambda^t_i := \# \{ j | \lambda_j \ge i \}
\label{transpdef}
\eeq
Note, that applying the conjugation twice returns the original partition $\lambda$.

Regardless of the amount of supersymmetry, there are several ways to define surface operators in gauge theories.
Thus, one can introduce a surface operator supported on $S$ by postulating a singularity for gauge fields along $S$ or,
alternatively, one may couple 4d gauge theory to a 2d theory $\TS$
with global (flavor) symmetry group $G$ \cite{Gukov:2006jk,Gukov:2008sn}.
Note, these two descriptions may not be unrelated, since by integrating out 2d degrees of freedom on $S$
one effectively generates source terms of the form $\int_S d^2 x (\ldots)$ in the Lagrangian of the four-dimensional theory.
Also, often there are multiple different choices of the 2d theory $\TS$ on $S$ that lead to the same surface operator;
typical examples include 2d theories of the previous section that flow to the same IR fixed point.

Although much of the discussion here related to the definition and classification of Levi types
applies to 4d gauge theories with $\CN=2$ as well as $\CN=4$ supersymmetry (and some even to $\CN=0$),
it is convenient to start with the larger amount of supersymmetry in order to have better control of
quantum effects and then discuss modifications due to lower amount of SUSY.
With this plan in mind, we start with surface operators in ${\cal N}=4$ super Yang-Mills
defined as singularities in the field configuration.

The field content of ${\cal N}=4$ SYM can be described conveniently in terms of ${\cal N}=1$ superfields: an ${\cal N}=1$ vector multiplet $W_{\alpha}$ and three $ $${\cal N}=1$ chiral multiplets $\Phi_{1},\Phi_{2}$ and $\Phi_{3}$. Superfields $W_{\alpha}$ and $\Phi_{1}$ combine to form an ${\cal N}=2$ vector multiplet, while $\Phi_{2}$ and $\Phi_{3}^{\dagger}$ combine to form an ${\cal N}=2$ hyper-multiplet. All the fields are in the adjoint representation of the gauge group $G$. Since the BPS equations are local, without loss of generality we can assume the support of the surface operator, $S$, to be oriented along the $(x^0,x^1)$ directions. Then, the BPS equations reduced to $(x^2,x^3)$ plane are,
\be \label{BPSeqs}
F_{23}-[\phi_{2},\phi_{2}^{\dagger}]  =  0,\qquad
D_{\bar{z}}\phi_{2}  =  0.
\ee
Here, $\phi_{2}$ is the scalar component of $\Phi_{2}$ and $z=x^{2}+ix^{3}$.
Perhaps a more familiar form of these equations is obtained if one considers
the geometric Langlands type twist of ${\cal N}=4$ theory where the $SO(4)$ holonomy
of the space is identified with the $SO(4)\subset SO(6)$  R-symmetry \cite{Marcus:1995mq,Kapustin:2006pk}.
The fields $\phi_{1}$ and $\phi_{2}$ change to a one-form $\phi_{\mu}dx^{\mu}$ with
$\phi_{\mu}=(\mbox{Re}\phi_{1},\mbox{Im}\phi_{1},\mbox{Re}\phi_{2},\mbox{Im}\phi_{2})$ and the equations \eqref{BPSeqs}
become\footnote{In fact, these equations belong to a one-parameter family of BPS equations
introduced in \cite{Kapustin:2006pk}. For the special values of the deformation parameter $t = \pm i$
these more general BPS equations reduce to \eqref{BPSmkw}, which is the only case we need for the present discussion.}
\be
F-\phi\wedge\phi  =  0,\qquad
d_{A}\phi  =  0,\qquad
d_{A}*\phi  =  0.
\label{BPSmkw}
\ee
In this form the BPS equations, restricted to the $(x^2,x^3)$ plane, are known as Hitchin's equations \cite{Hitchin:1986vp}.
The surface operator is introduced by requiring the fields to satisfy the BPS condition
everywhere except at its support i.e. at the origin of the $(x^2,x^3)$ plane.
The simplest of such solutions is the one where $\phi$ has a simple pole at $r=0$,
\be\label{singularsol}
A  =  \alpha d\theta+\ldots, \qquad\qquad \phi  =  \beta\frac{dr}{r}-\gamma d\theta+\ldots.
\ee
The constants $\alpha,\beta$ and $\gamma$ are valued in the part of the Lie algebra of $G$ invariant
under the Levi subgroup $\bbL$, and the ellipses denote the regular part of the solution.
Having only a simple pole makes this singularity a ``tame'' singularity, as opposed to the ``wild'' singularity which corresponds to higher order poles.
The surface operator defined by the tame singularity is called the tame surface operator,
which will be our prime class of examples in this paper.

In addition to the $\mathbb{L}$-invariant constants $(\alpha,\beta,\gamma)$ parametrizing the singularity,
we add a term $\eta\int_{S}F$ to the action where $S$ denotes the support of surface operator.
Then, in the $\mathbb{L}$-invariant parameters $(\alpha,\eta,\beta,\gamma)$ are known as the defining parameters of the surface operator.
This surface operator preserves $\CN = (4,4)$ supersymmetry from the point of view of unbroken 2d Lorentz symmetry.
As we mentioned earlier, in 4d theories with $\CN=2$ supersymmetry the $\CN=4$ super-multiplet
splits into a vector multiplet and a hypermultiplet,
so that the gauge field $A$ and the Higgs field $\phi$ are no longer in the same multiplet
(and, in fact, the field $\phi$ is no longer required to be in the adjoint representation of the gauge group).
This leads to various generalizations of the BPS equations \eqref{BPSmkw} in $\CN=2$ gauge theories
(see e.g. \cite[sec.3]{Gukov:2007ck} and, in particular, section \ref{N2} below).

Note, the Levi subgroup serves as the only discrete label of such surface operator.
Since the main purpose of this paper is to study the superconformal index of ${\cal N}=2$ theories with surface operators
and the index does not depend on any continuous parameters of the theory,
the exact values of $(\alpha,\eta,\beta,\gamma)$ will play a secondary role in our story.
We will only use the discrete label $\mathbb{L}$ to classify the surface operators.

\subsection*{$2d$-$4d$ system}

Following \cite{Gukov:2006jk,Gukov:2008sn} we would like to describe the surface operator as a coupled $2d$-$4d$ system.
It involves introducing new degrees of freedom on the support of the surface operator and coupling them to the bulk $4d$ gauge theory.
If the $2d$ theory satisfies certain criteria then integrating it out completely introduces the desirable singularity
in four dimensional bulk fields. A familiar example of such $2d$ theory is a non-linear sigma model,
whose target space should satisfy the following conditions to give a half-BPS surface operator in $\CN=4$ SYM:
\begin{itemize}
\item It should be hyper-K\"ahler so that it preserves $(4,4)$ supersymmetry.
\item It should admit a $G$ action so that it can be coupled to $4d$ gauge symmetry.
\item It should be labelled by the Levi subgroup $\bbL$.
\end{itemize}
In the case of $\CN=2$ gauge theory that will be discussed later, the first condition is relaxed to manifolds with K\"ahler structure,
whereas in $\CN=0$ case the analog of such surface operators is obtained by keeping only two last conditions.
In the $\CN=4$ case, the space $T^*(G/\bbL)$ satisfies all the criteria and provides an ideal candidate for the target
manifold\footnote{Note that the moduli space of the solutions with prescribed singularity \eqref{singularsol} of type $\bbL$ is $T^{*}(G/\mathbb{L})$.}.
The coset $G/\mathbb{L}$ is a coadjoint orbit for the group $G$ and, as a K\"ahler manifold, is a good starting point
for constructing surface operators in $\CN=2$ gauge theory.

To see the relation of this approach to \eqref{singularsol}, let us take a simple example with $G=SU(2)$ and $\mathbb{L}=U(1)$. In this case, $G/\bbL$ is simply $\bbCP^1$. The parameters $(\alpha,\beta,\gamma)$ of the surface operators are encoded in the moduli of $T^*\bbCP^1$, and the parameter $\eta$ is the $B$-field through the $\bbCP^1$.
\begin{center}
\begin{tabular}{|c|c|c|}
\hline
Complex structure & K\"ahler modulus & Complex modulus\tabularnewline
\hline
\hline
$I$ & $\alpha$ & $\beta+i\gamma$\tabularnewline
\hline
$J$ & $\beta$ & $\gamma+i\alpha$\tabularnewline
\hline
$ $$K$ & $\gamma$ & $\alpha+i\beta$\tabularnewline
\hline
\end{tabular}
\par\end{center}

\begin{figure}
\centering
\includegraphics[scale=0.4]{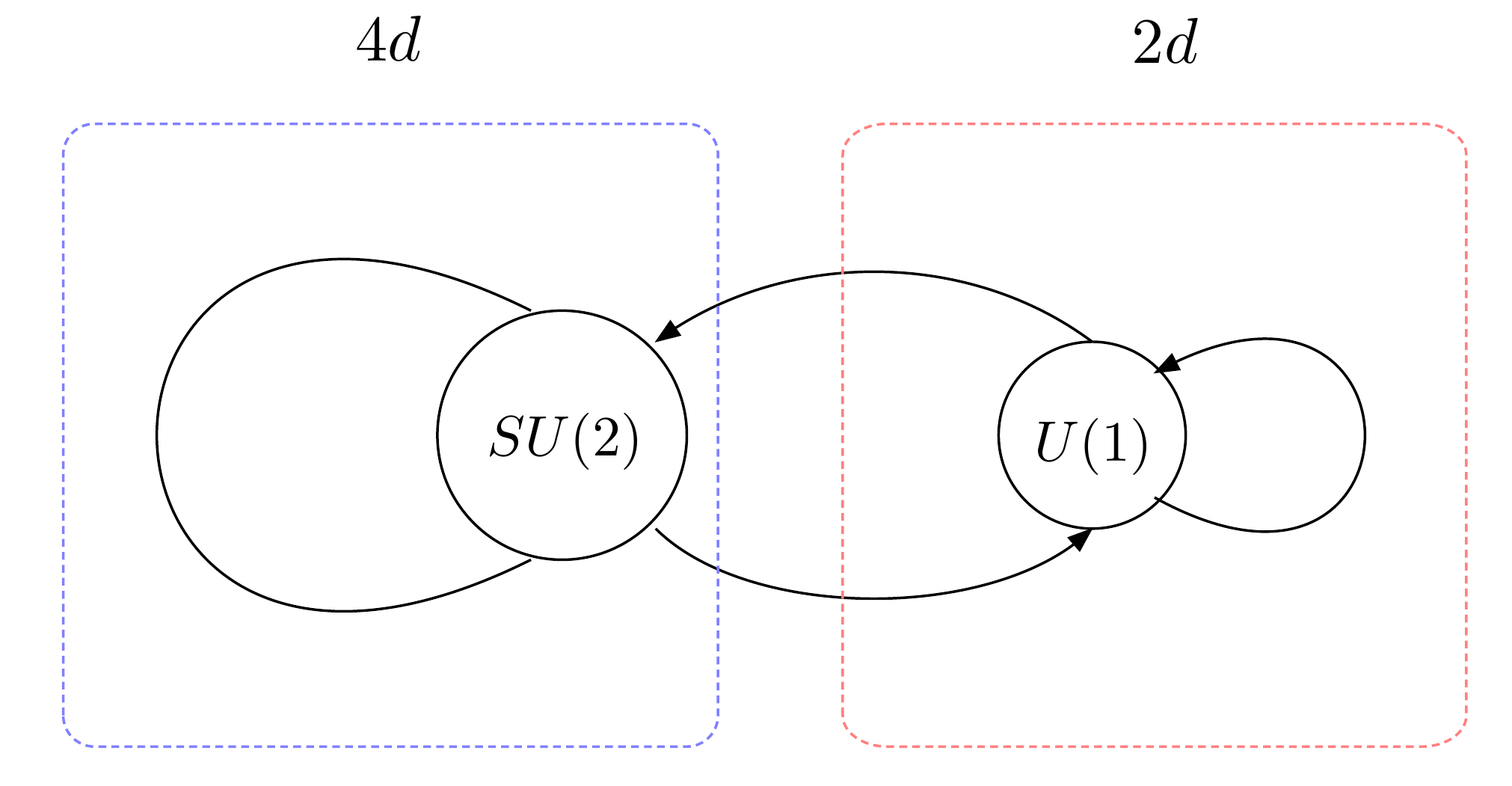}
\caption{$SU(2)$ ${\cal N}=4$ SYM coupled to the $U(1)$ gauge theory living on the support of the surface operator. The $4d$ field content is denoted in terms of ${\cal N}=2$ multiplets while the $2d$ field content is denoted in terms of $(2,2)$ multiplets.}
\label{2d4dquiver}
\end{figure}

In order to compute the superconformal index, we prefer to work with the gauge theory description. In this example, the target $T^*\bbCP^1$ can be constructed as the moduli space of the $U(1)$ gauge theory with two hypermultiplets \emph{i.e.} two chiral multiplets $q_{i}$ with charge $+1$ and two chiral multiplets ${\tilde q}_i$ with charge $-1$. The K\"ahler modulus of the sigma model is the FI parameter of the $2d$ gauge theory. It combines with the $2d$ theta angle $\eta$ to form a complex FI parameter $t:=\alpha+i\eta$. The complex structure modulus $\beta+i\gamma$ is encoded in the $2d$ superpotential. This coupled system of $4d$ and $2d$ gauge theory is conveniently summarized in the quiver diagram \ref{2d4dquiver}.

The BPS equations \eqref{BPSeqs} of the $4d$ ${\cal N}=4$ gauge theory are modified due to the coupled $2d$ theory. The $2d$ contribution is only supported at the origin of $(x^2,x^3)$ plane:
\begin{eqnarray}\label{2d4dBPS}
F_{23}-[\phi_{2},\phi_{2}^{\dagger}] & = & \delta^{(2)}qq^{\dagger}\nonumber\\
D_{\bar{z}}\phi_{2} & = & \delta^{(2)}q\tilde{q}
\end{eqnarray}
The first equation follows from the D term while the second is the equation of motion coming from a novel superpotential term $W=\mbox{tr}\,\phi_2 D_{\bar z} \phi_2$ in $4d$. Although the $4d$ supersymmetry doesn't allow it, the $(4,4)$ supersymmetry algebra in $(x^0,x^1)$ directions which is the only preserved supersymmetry algebra, does allow this term \cite{Constable:2002xt}. Using the BPS equations of the $2d$ theory, the bilinears $qq^{\dagger}$ and $q\tilde{q}$ can be eliminated in favor of  the FI parameter $\alpha+i\eta$ and the superpotential parameter $\beta+i\gamma$. The $\delta$ function source in \eqref{2d4dBPS} induces the singularity \eqref{singularsol} in the solution. This example shows the relation between the two approaches to $4d$ surface operators: first as a prescribed singularity and second as a $2d$-$4d$ coupled system. For us, the $2d$-$4d$ quiver is the most convenient description of the surface operator. As will be shown explicitly in the later part of the paper, describing any surface operator in this way allows us to straightforwardly compute its superconformal index. We proceed to derive the gauge theory that engineers the sigma model on $T^*(G/\bbL)$ corresponding to the surface operators of general Levi type  $\bbL$. We denote this gauge theory as $\TS$.

\subsection*{$2d$ gauge theory for surface operator of type $\bbL$}

As we already explained in \eqref{Levilambda}, a general Levi subgroup of $G=SU(N)$ is labeled by
a partition of $N$ and has the form $\bbL = S[U(k_1)\times \ldots \times U(k_n)]$,
where $k_i$ are the parts of the conjugate partition \eqref{laprel}.
We would like to construct a $2d$ $\CN=(4,4)$ gauge theory which engineers a non-linear sigma model on $X=T^*(G/\bbL)=G_{\bbC}/\bbL_\bbC$. As pointed out earlier, $X$ admits hyper-K\"ahler metric and enjoys a $G$ action so they provide ideal candidates for the target manifolds of half-BPS surface operators in $\CN=4$ SYM. From \eqref{Levilambda} it is clear that the complex dimension of $X$ is
\be
\mbox{dim}_\bbC X=N^2-\sum_{i=1}^n k_i^2
\label{CMdim}
\ee
which is the familiar formula for the dimension of the conjugacy class labeled by the partition $\lambda$ (see \cite{CMbook}, section 6.1).

In order to obtain a gauge theory description, we need to describe $X$ as a hyper-K\"ahler quotient of a vector space $V$ by a group $G_{2d}$ that will be interpreted as the gauge group for the $2d$ theory $\TS$. For surface operators of Levi type $\mathbb{L}$, there is indeed a way to represent the complex conjugacy class $G_\bbC/\bbL_\bbC$ as a hyper-K\"ahler quotient \cite{KSwann}, which was already used in the study of half-BPS surface operators \cite{Gukov:2006jk} and\footnote{In $\CN=4$ gauge theory, the supersymmetry equations for both half-BPS surface operators and boundary conditions both reduce to Nahm equations resulting in a natural bijection between the two classes of objects.} boundary conditions in $\CN=4$ super-Yang-Mills theory \cite{Gaiotto:2008sa}. Interested in the former application, here we review the construction of \cite{KSwann}.

Consider a two-dimensional theory $\TS$ defined by a linear quiver with $n$ nodes,
the last of which corresponds to a non-dynamical flavor symmetry group $G=SU(N)$:

\begin{figure}
\centering
\includegraphics[scale=0.5]{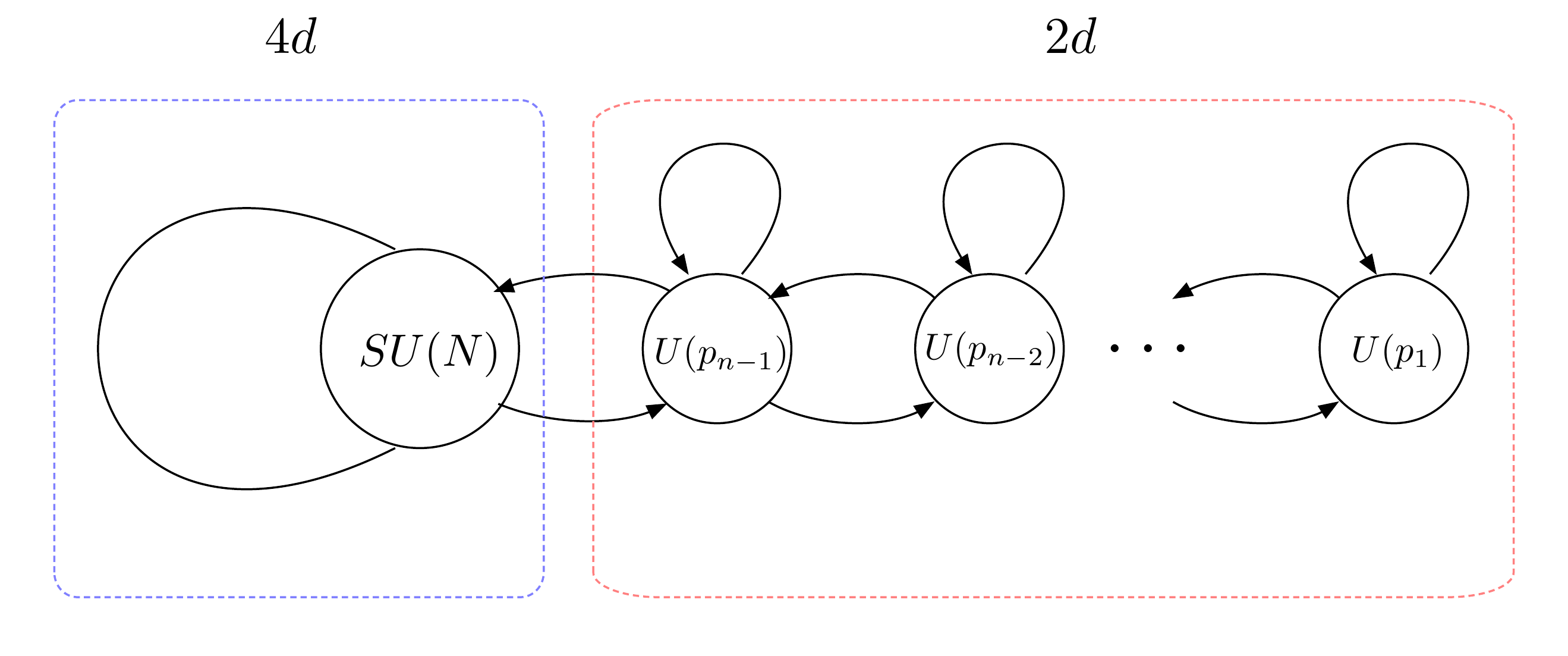}
\caption{The theory $\TS$ is depicted in the red dotted box (on the right hand side). We have used $(2,2)$ multiplets to denote the field content of this $(4,4)$ theory. The $SU(N)$ flavor symmetry of $\TS$ is coupled to the $4d$ ${\cal N}=4$ SYM (in blue dotted box) as shown.}
\label{N4general}
\end{figure}

The flavor symmetry $SU(N)$ is used to couple the $2d$ theory to the $4d$ theory. The field content of $\TS$ is,
$$
\TS  =
\begin{cases}
& \text{2d theory with gauge group}~ G_{2d} = U(p_1) \times U(p_2) \times \ldots \times U(p_{n-1}), \\
& \text{bi-fundamental hypers}~ {\bf (p_i, p_{i+1})} ~\text{and}~ N ~\text{fundamental hypers for}~ U(p_{n-1})
\end{cases}
$$
with $p_i < p_{i+1}$. The Higgs branch of this theory is hyper-K\"ahler quotient $Y = V /\!/\!/ G_{2d}$, where $V$ is a vector space spanned by the hypermultiplets. It has the quaternionic dimension
\beq
\dim_{\mathbb{H}} V \; = \; p_1 p_2 + p_2 p_3 + \ldots + p_{n-1} N.
\eeq
According to \cite{KSwann}, the hyper-K\"ahler quotient $Y$ is (the closure of) the nilpotent orbit $\CO = G_{\C} \cdot x$ of an element $x \in \mathfrak{g}_{\C}$,
such that
\beq
\text{rank} (x^i) = p_{n-i}.
\label{ranknki}
\eeq
For unitary groups, all nilpotent orbits are Richardson, meaning that they can be deformed to semi-simple orbits
and -- via the exponential map -- identified with conjugacy classes $G_{\C} / \mathbb{L}_{\C}$, see {\it e.g.} \cite{Gukov:2006jk}.
Therefore, modulo details related to the center and topology of the gauge group $G$ (which play absolutely no role in this paper),
we can write
\beq
\exp: \quad Y=\CO ~ \xrightarrow[~]{~\cong~} ~ G_{\C} / \mathbb{L}_{\C}=X.
\label{expmap}
\eeq
Moreover, the relation \eqref{ranknki} makes it clear that the number $p_i$
is equal to the sum of $k_j$ with $j \le i$:
\beq\label{pkrelation}
p_i \; = \; \sum_{j=1}^i k_j,\qquad \mbox{i.e.}\qquad k_i = p_i - p_{i-1}
\eeq
with $p_n = N$ and $p_0 = 0$. Using this formula, we can verify that indeed $\mbox{dim}_\bbC Y=\mbox{dim}_\bbC X$:
\beq
\mbox{dim}_\bbC Y=2 \sum_{i=1}^{n-1} p_i p_{i+1} - 2 \sum_{i=1}^{n-1} p_i^2 \; = \;  N^2 - \sum_{i=1}^n k_i^2=\mbox{dim}_\bbC X.
\eeq

\begin{figure}
\centering
\includegraphics[scale=0.4]{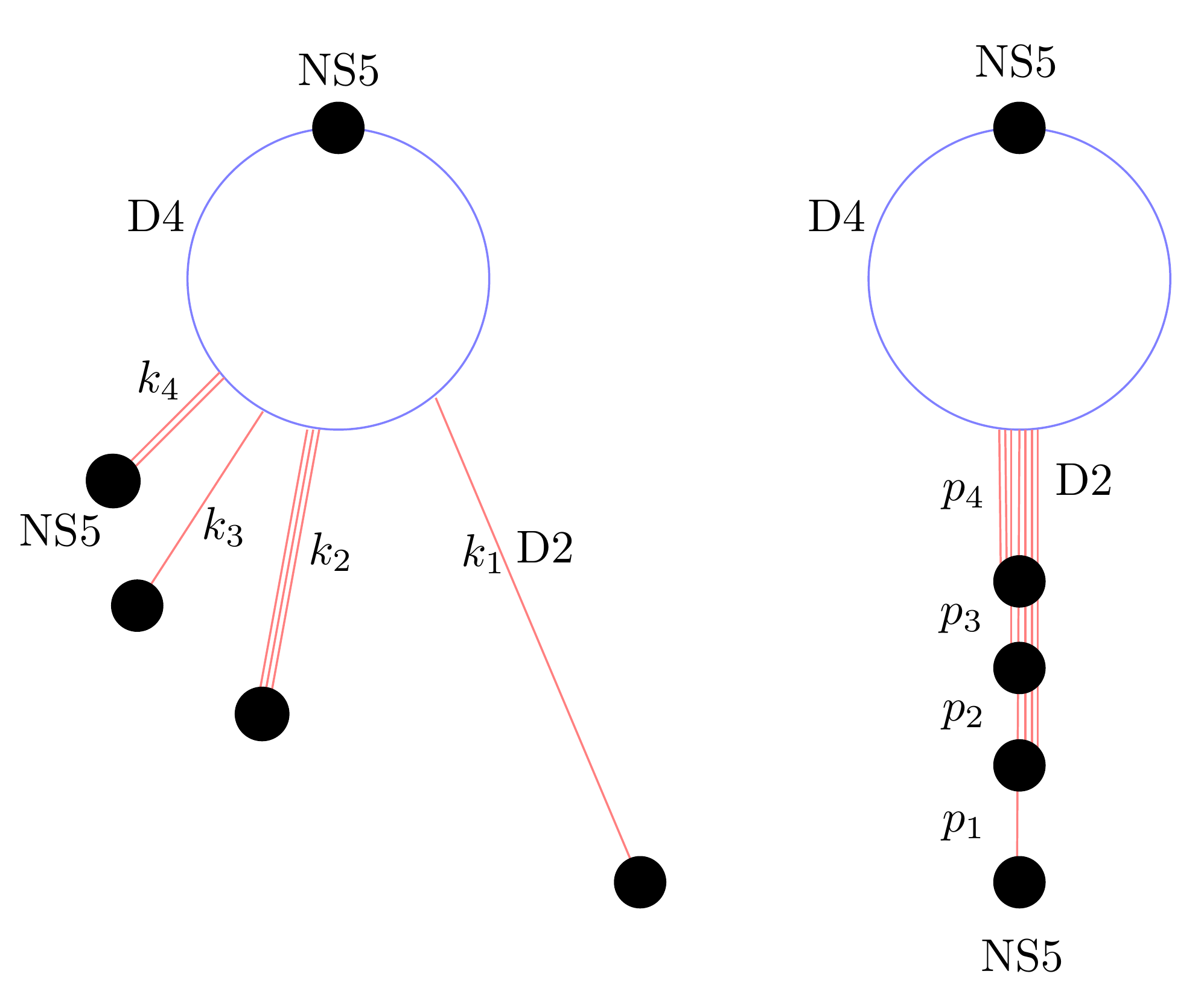}
\caption{The brane set up realizing 2d/4d coupling. In this example, we have chosen $n=5$.}
\label{N4branes}
\end{figure}

The two-dimensional theory $\CT_{2d}$ with gauge group $G_{2d} = \prod_{i=1}^{n-1} U(p_i)$ can be conveniently
realized on the world-volume of D2-branes stretched between $N$ D4-branes and $(n-1)$ NS5$'$-branes.
By placing NS5$'$-branes in a generic position (so that no two are aligned in the $x^6$ direction),
it is easy to infer from eq. \eqref{pkrelation} that $k_i$ is simply the number of D2's stretched between
the $i$-th NS5$'$-brane and $N$ D4-branes.
Then, placing all NS5$'$ and D2 branes in the same position along the $x^6$ direction yields a brane
realization of the two-dimensional theory $\CT_{2d}$ with gauge group $G_{2d} = \prod_{i=1}^{n-1} U(p_i)$. This brane set up is depicted in figure \ref{N4branes}.

The mathematical structure of the two-dimensional theory $\TS$ encountered here is best described by the so-called {\it flag} $F$,
that is a sequence of subspaces $(F_0, \ldots, F_n)$, such that $\dim F_i = p_i$ and
\beq
\{ 0 \} = F_0 \subset F_1 \subset F_2 \subset \ldots \subset F_{n-1} \subset F_n = \mathbb{C}^N
\eeq
with $F_{i-1} \ne F_i$ for $1 \le i \le n$. Note, that according to \eqref{pkrelation}, $k_i = \dim (F_i / F_{i-1})$. The stabilizer of the flag $F$ is the parabolic subgroup $\CP$ of $G_{\C} = SL(N,\C)$ whose Levi subgroup $\mathbb{L} = G \cap \CP$ is precisely $S[U(k_1)\times\ldots\times U(k_n)]$, as claimed in \eqref{Levilambda}. The space of such flags is a quotient space,
\beq
G/\mathbb{L} \; \cong \;  G_{\C} / \CP \,,
\eeq
called the {\it partial flag variety}.
The Poincar\'e polynomial of the partial flag variety is given by the following well-known formula:
\beq
P (G/\bbL;t) \; = \; \sum_{i} b_{i} (G/\bbL) t^i \; = \;
\frac{\prod_{i=1}^N (1-t^{2i})}{\prod_{j=1}^{n} \prod_{i=1}^{k_j} (1-t^{2i})}.
\label{flagcohom}
\eeq
Note that $P(X;t)=P(G/\bbL)$ since the space $X$ retracts to $G/\mathbb{L}$. The sequence $(k_1, \ldots, k_n)$ is called the type of the flag $F$ (and of the corresponding parabolic subgroup $\CP$). Parabolic subgroups are conjugate if and only if they have the same type.
On the other hand, in agreement with the above discussion,
Levi factors of the parabolic subgroups are labeled by partitions $\lambda$ or their conjugates \eqref{laprel}.
In other words, $\lambda^t$ is a non-increasing sequence obtained by a permutation of $(k_1, \ldots, k_n)$ and
$\lambda_i \; = \; \# \{ j \vert k_j \ge i \}$.
Two parabolic subgroups with flag types $(k_1, \ldots, k_n)$ and $(k_1', \ldots, k_n')$ have conjugate Levi factors if and only if
$\text{ord} \{k_1, \ldots, k_n\} \; = \; \text{ord} \{k_1', \ldots, k_n'\}$. This means we can use $k'_i$ instead of $k_i$ in \eqref{pkrelation} to obtain the gauge theory $\TS'$. The theories $\TS$ and $\TS'$ are expected to be dual to each other and either can be used to construct the surface operator of type $\bbL$.
Note, that exchanging the order of the NS5-branes in figure \ref{N4branes} corresponds to different choices of the parabolic subgroup $\CP$ (associated to a given Levi $\bbL$). The process of relating two brane configurations with different arrangements of $(p_1,\ldots, p_n)$ D2-branes can be understood as carrying the NS5-branes around each other or, if they are aligned, by passing them through each other. In the latter case, the brane creation mechanism encountered here is essentially the familiar brane realization of the Seiberg duality in $4d$ $\CN=1$ gauge theories \cite{Elitzur:1997hc}.

Conversely, given a partition $\lambda$ which labels the Levi type,
the number of parabolic subalgebras with this Levi factor is given by
\beq
N_{\text{par}} (\lambda) \; = \; \frac{\lambda_1 !}{\prod_{i \ge 1} (\lambda_i - \lambda_{i+1})! }
\label{Nparfla}
\eeq
Thus, for $G = SU(4)$ we have 5 different Levi types
which correspond to five conjugacy classes in $SL(4,\C)$ labeled by partitions of 4:
\vskip 1em \centerline{\vbox{\halign{\quad # & \quad\quad\quad # \cr
$\underline{~~~~\lambda~~~~}$ & $\underline{N_{\text{par}} (\lambda)}$
\cr
$[4]$   & $1$ \cr
$[3,1]$   & $3$  \cr
$[2,2]$   & $1$  \cr
$[2,1,1]$ & $2$  \cr
$[1,1,1,1]$ & $1$ \cr }}}
\noindent
In total, in this case one finds 8 conjugacy classes of parabolic subalgebras, in agreement with \eqref{Nparfla}.
For example, there are three parabolic subalgebras of type $(2,1,1)$, $(1,2,1)$ and $(1,1,2)$
associated with the Levi type \eqref{Leviex} indexed by the partition $\lambda = [3,1]$.
Using \eqref{flagcohom}, one can easily find cohomology of the corresponding partial flag variety:
\beq
X = \frac{SU(4)}{S(U(2) \times U(1) \times U(1))} : \qquad P(X;t) \; = \; 1 + 2t^2 + 3t^4 + 3t^6 + 2t^8 + t^{10}
\eeq
Note, this partial flag variety has complex dimension 5, in agreement with \eqref{CMdim}.

\subsection*{Examples}

Let us illustrate this construction with examples.
Consider $SU(N)$ gauge theory with a surface operator of Levi type $\bbL=S[U(k)\times U(N-k)]$
labeled by the partition $\lambda = [2^k,1^{N-2k}]$ (where $n \le N-n$ is assumed).
It is easy to check that, in this case, the parts of the conjugate partition are $(k_1,k_2) = (k,N-k)$ or $(N-k,k)$.
The $\TS$ gauge theory is given by a single node $U(k)$ or $U(N-k)$ depending upon whether we take
the sequence corresponding to the parabolic subgroup to be $(k,N-k)$ or $(N-k,k)$, and is realized on a single stack of D2-branes.
In either case, the Higgs branch of such theory is the cotangent bundle of Grassmannian $\mbox{Gr}(k,N)$.
The Grassmannian is isomorphic to the quotient,
\be
\mbox{Gr}(k,N)\simeq SU(N)/S[U(k)\times U(N-k)].
\label{Grquot}
\ee
As it is manifestly symmetric under the exchange $k\leftrightarrow N-k$, $U(k)$ and $U(N-k)$ gauge theories are dual to each other.
This statement can be thought of as the $(4,4)$ supersymmetric version of the Hori-Tong duality.
In fact, successive application this duality can be used to show that the gauge theories
resulting from different parabolic subgroups but same Levi subgroup are dual to each other. As pointed out earlier, in the brane setup, these dualities are realized as shuffling the order of NS5 branes in figure \ref{N4branes}.

Further specialization to $k=1$ (or $k=N-1$) gives $X = T^* \C {\bf P}^{N-1}$,
whereas specialization to $k=0$ gives the trivial orbit $X = \{ 0 \}$ labeled by $\lambda = [1^N]$.
The other extreme case, $\lambda=[N]$, corresponds to $\mathbb{L} = \mathbb{T}$
and gives a {\it regular} orbit $\CO_{\text{reg}}$ of maximal dimension.
(For more general gauge groups, the dimension of this maximal orbit is equal to $\dim G - \text{rank} \, G$.)

\subsection{Surface operators in ${\cal N}=2$ SCFTs}
\label{N2}

In this section we discuss the half-BPS surface operators in $\CN=2$ superconformal gauge theories of quiver type.
This allows a fairly general class of theories, with some nodes of the quiver representing gauge symmetries
and others global flavor symmetries, which are not gauged.
The basic building block of such quiver theory --- that can be viewed as a 4d $\CN=2$ SCFT on its own ---
is $SU(N)_G$ $\CN=2$ gauge theory with $2N$ flavors also known as the $\CN=2$ SCQCD.
\begin{figure}[h]
\centering
\includegraphics[scale=0.4]{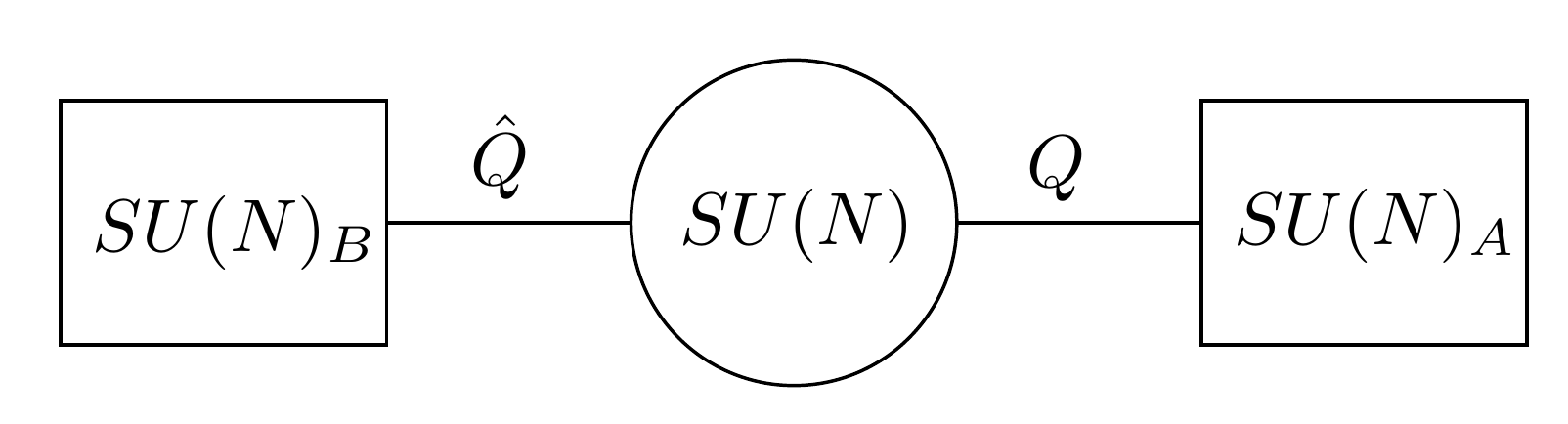}
\caption{A convenient quiver representation of the $\CN=2$ SCQCD.}
\label{N2sqcd}
\end{figure}
It is convenient to split the $2N$ hypermultiplets of SCQCD into two sets of $N$ hypermultiplets each, $Q_i$ transforming in the fundamental of $SU(N)_A$ and ${\hat Q}_i$ transforming in the fundamental of $SU(N)_B$, see figure \ref{N2sqcd}.
As in the previous discussion, we take the surface operator to be oriented along $(x^0,x^1)$ directions.
The BPS equations reduced to the transverse $(x^2,x^3)$ plane are similar to the equations \eqref{BPSeqs}:
\be\label{N2surface}
{F_{23}}_i^j - Q_i^k\, {Q^\dagger}_k^j= 0,\qquad \qquad {D_{\bar z}\,}_i^j Q_j^k= 0.
\ee
The surface operator is introduced by having the fields satisfy the BPS condition everywhere except at its support i.e. at the origin of the $(x^2,x^3)$ plane. Analogous to the surface operators of $\CN=4$ theory, the surface operators of the $\CN=2$ theory can also be constructed by coupling it to a $2d$ gauge theory $\TS$.

In this case, $\TS$ is a 2d gauge theory with $\CN=(2,2)$ supersymmetry. Unlike for $\CN=4$ SYM, its flavor symmetry could either be $SU(N)$ or $SU(N)\times SU(N)$. In the former case, it is coupled to the $4d$ theory by identifying the $SU(N)$ flavor symmetry with the gauge symmetry $SU(N)_G$ of the $4d$ theory while in the later case, the coupling involves identifying one $SU(N)$ factor with $SU(N)_G$ and the other with one of the flavor symmetry factors of the SCQCD, say $SU(N)_A$. It turns out that the index of the surface operators of the later case is invariant under four dimensional S duality. We will focus on that case from now on. To summarize, the criteria $\TS$ should satisfy are:
\begin{itemize}
\item It should be $\CN=(2,2)$ supersymmetric.
\item It should have $SU(N)\times SU(N)$ flavor symmetry.
\item It should be labelled by the Levi subgroup $\bbL$.
\end{itemize}
The third criterion is necessary only if one want to construct surface operators that correspond
to tame singularity in the field configuration analogous to those in $\CN=4$ SYM.
In section \ref{vortex} we will relax this criterion but for now we will stick with it.
The simplest theory satisfying the first two criteria is the $U(k)$ gauge theory with $N$ fundamental and $N$ anti-fundamental chiral multiplets studied in the section \ref{nonabelian-without}. As illustrated in that section, this theory enjoys a Hori-Tong type duality under the exchange of $k\leftrightarrow N-k$. This strongly suggests that it should correspond to the surface operator of type $\bbL=S[U(k)\times U(N-k)]$ because this Levi subgroup is also symmetric under the exchange of $k$ and $N-k$ as well. In fact, this leads to a natural conjecture for the surface operator of type $\bbL$:

A simple heuristic way to guess the spectrum of the 2d theory $\TS$ in the $\CN=2$ case
could be based on the general remark made in the beginning of section \ref{N4} that half
of the supermultiplet content is projected out upon reduction from $\CN=4$ to $\CN=2$ SUSY.
Therefore, one might expect that, for the same choice of symmetries (such as $G$ and $\bbL$),
the spectrum of the 2d theory $\TS$ in $\CN=2$ case is roughly half of that in the $\CN=4$ SYM.
This reduction can be described geometrically by replacing quaternionic spaces $\mathbb{H}^r$
(parametrized by 2d hypermultiplets) with complex spaces of the same dimension $\C^r$
(parametrized by 2d chiral multiplets) and by replacing hyper-K\"ahler quotients $\mathbb{H}^r /\!/\!/ G_{2d}$
with the K\"ahler quotients $\C^r /\!/ G_{2d}$ (that besides gauge conjugation implement only D-term, but not F-term constraints).

\begin{figure}[h]
\centering
\includegraphics[scale=0.4]{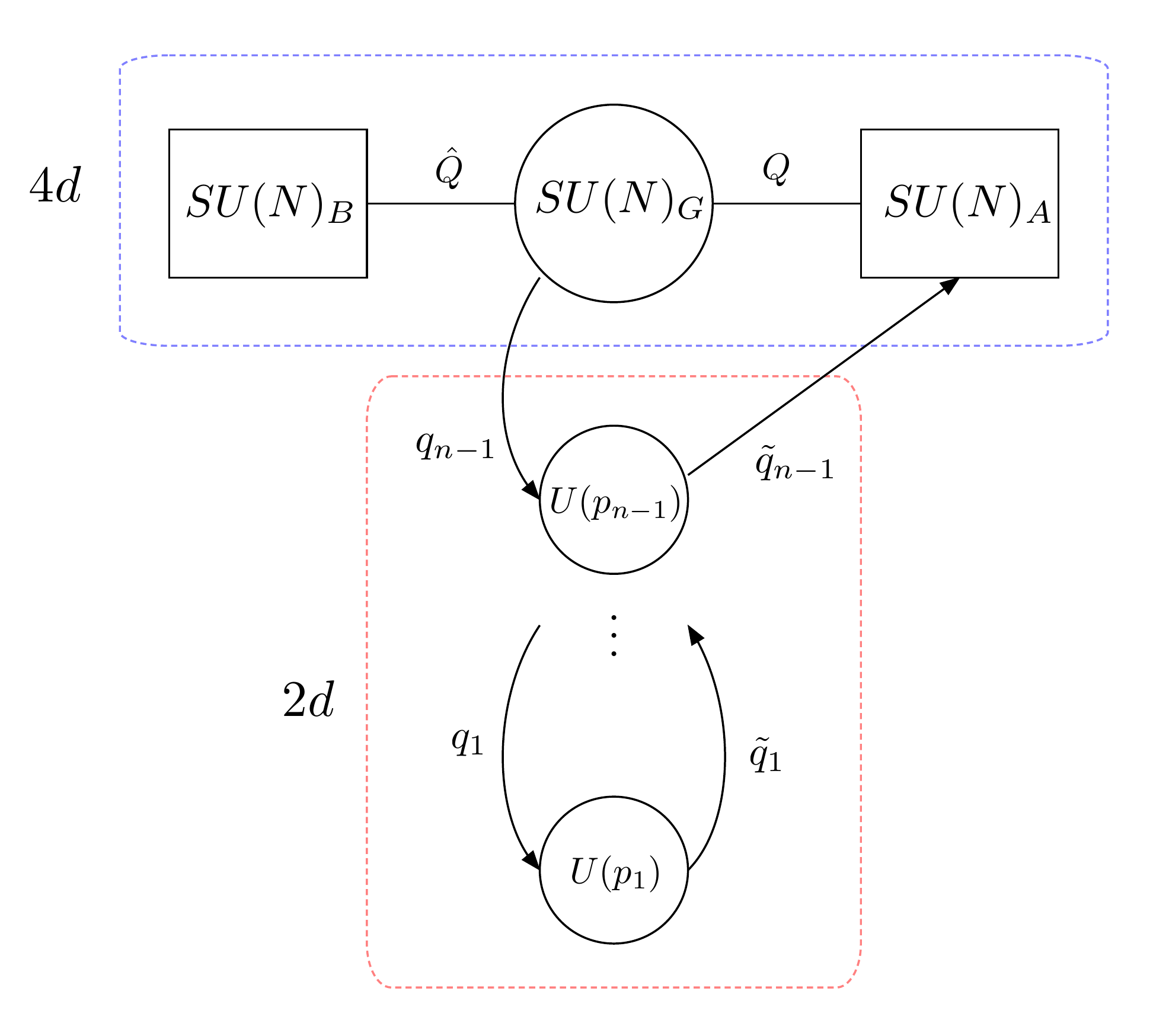}
\caption{The quiver diagram describing the $2d$-$4d$ system for the surface operator of Levi type $\bbL=S[U(k_1)\times \ldots \times U(k_n)]$. The ranks of the $2d$ gauge group $p_i=\sum_{j=1}^{i} k_j$.}
\label{N2general}
\end{figure}

For example, in the simple case of the Grassmannian sigma-model \eqref{Grquot}, this reasoning
leads precisely to the $U(k)$ Hori-Tong theory with $N$ fundamental chiral multiplets and the Higgs branch $\C^{kN} /\!/ U(k)$.
However, as we mentioned earlier, the reduction from $\CN=4$ to $\CN=2$ should be also accompanied by
``doubling'' the symmetry group $SU(N) \to SU(N) \times SU(N)$ of the 2d theory $\TS$.
This effectively puts the removed chiral multiplets back in: $\C^{kN} \to \C^{kN} \oplus \C^{kN}$.
Motivated by this, one natural guess for the 2d theory $\TS$ associated to a surface operator
of Levi type $\bbL$ is
$$
\TS  =
\begin{cases}
& \text{2d theory with gauge group}~ G_{2d} = U(p_1) \times U(p_2) \times \ldots \times U(p_{n-1}), \\
& \text{bi-fundamental hypers}~ {\bf (p_i, p_{i+1})} ~\text{and}~ N ~\text{fundamental hypers for}~ U(p_{n-1})
\end{cases}
$$
where each bi-fundamental hyper is now regarded as a (fund, anti-fund) pair of chirals (going in opposite ways),
and the gauge theory nodes only contain $\CN=2$ vector multiplets (i.e. without extra adjoints).
The last set of $N$ fundamental hypers for $U(p_{n-1})$ is, therefore, interpreted as
a pair of bifundamental chirals $({\bf p_{n-1}}, {\bf N}_G) \oplus ({\bf N}_A,{\bf p_{n-1}})$. The quiver diagram for the $2d$-$4d$ system is shown in figure \ref{N2general}. Note, since the numbers of fundamental and anti-fundamental chiral multiplets are equal at each gauge node, this guarantees cancellation of anomalies, which of course is required for superconformal invariance in two dimensions.
In the remainder of the section we will compute the superconformal index of the half-BPS surface operator with $\bbL=S[U(k)\times U(N-k)]$. It amounts to setting $n=2$ and $p_1=k$ in the quiver diagram of figure \ref{N2general}.

\subsection{Index of the surface operator}\label{part3}

In order to compute the superconformal index of the $2d$-$4d$ coupled system we need to describe the embedding of $2d$ $(2,2)$ superconformal algebra into $4d$ $\CN=2$ superconformal algebra. This embedding related the fugacities $(q,t)$ used in computing the $2d$ $(2,2)$ index with the fugacities $(\fp,\fq,\ft)$ used in computing the $4d$ $\CN=2$ index. Let us start with a brief review of the $4d$ index.

\subsection*{4d index}

The four dimensional index is also a powerful quantity, encodes the superconformal spectrum and can be computed in weak coupling limit. It has been used to check Gaiotto's conjectures \cite{Gaiotto:2009we} of S-duality in ${\cal N}=2$ theory \cite{Gadde:2011uv}.
Let $E$ be the conformal dimension and $h_{01}$ and $h_{23}$ be the rotation generators in the $01$ and $23$ planes, respectively. They are related to the Cartans $j_1,j_2$ of $SU(2)_1\times SU(2)_2$ as, $h_{01}=j_1+j_2$ and $h_{23}=-j_1+j_2$.
Letting $R$ be the Cartan of $SU(2)_R$ symmetry and $r$ be the $U(1)_r$ charge, the superconformal index for ${\cal N}=2$ theories is defined as
\be
{\cal I}_{4d}={\rm Tr} (-1)^F \fp^{h_{23}-r} \fq^{h_{01}-r} \ft^{R+r}.
\label{indexdef}
\ee
The fugacities $\fp$, $\fq$, and $\ft$ keep track of maximal set of quantum numbers commuting with a particular supercharge $\QQ$, which is chosen to be ${\tilde \QQ}^1_{\dot -}$  without loss of generality. It has $R=\frac{1}{2},\,r=-\frac{1}{2},\,h_{01}=-\frac{1}{2},\, h_{23}=-\frac{1}{2}$ and, of course, $E=\frac{1}{2}$. It is a simple matter to check that the charges appearing in the definition \eqref{indexdef} indeed commute with $\QQ$. Using the standard arguments for Witten index, only the states that obey $\delta\equiv 2\{\QQ,\QQ^\dagger\}=E-h_{01}-h_{23}-2R+r=0$ contribute to the superconformal index.

For a theory with weakly coupled Lagrangian description the index is computed by a matrix integral:
\be\label{4dcompute}
{\cal I}_{4d}(\fp,\fq,\ft)=\int [dU] \exp \Big( \sum_{n=1}^{\infty} \sum_{j}\frac{1}{n} f^{(j)_{4d}}(\fp^n,\fq^n,\ft^n) \chi_{R_j}(U^n,V^n)\Big).
\ee
Here, $U$ and $V$ denote elements of gauge and flavor groups, respectively. The invariant Haar measure integral $\int [dU]$ imposes the Gauss law over the Fock space. The sum is  over different ${\cal N}=2$ supermultiplets appearing in the Lagrangian,
with $R_j$ being the representation of the $j$-th multiplet under gauge and flavor group, and $\chi_{R_j}$ the character of $R_j$. The function $f^{(j)}$ is called single letter index. It is equal to either $f^V_{4d}$ or $f^{\frac{1}{2}H}_{4d}$ depending on whether the $j$-th multiplet is ${\cal N}=2$ vector multiplet or half-hypermultiplet. They are easily evaluated \cite{Gadde:2011uv} by listing contributions of the letters with $\delta=0$ in table \ref{single letters}:
\begin{table}
\begin{centering}
\begin{tabular}{|c|c|}
\hline
Letters with $\delta=0$ & index\tabularnewline
\hline
\hline
$\phi$ & $\fp\fq/\ft$\tabularnewline
\hline
$\lambda_{1\pm}$ & $-\fp,\,-\fq$\tabularnewline
\hline
$\bar{\lambda}_{1\dot{+}}$ & $-\ft$\tabularnewline
\hline
$\bar{F}_{\dot{+}\dot{+}}$ & $\fp\fq$\tabularnewline
\hline
$\partial_{-\dot{+}}\lambda_{1+}+\partial_{+\dot{+}}\lambda_{1-}=0$ & $\fp\fq$\tabularnewline
\hline
\hline
$\fq$ & $\sqrt{\ft}$\tabularnewline
\hline
$\bar{\psi}_{\dot{+}}$ & $-\fp\fq/\sqrt{\ft}$\tabularnewline
\hline
\hline
$\partial_{\pm\dot{+}}$ & $\fp,\, \fq$\tabularnewline
\hline
\end{tabular}
\par\end{centering}
\caption{Contributions to the index from ``single letters'' with $\delta=0$. We denote by $(\phi,\bar{\phi},\lambda_{I\alpha},\bar{\lambda}_{I\dot{\alpha}},F_{\alpha\beta},\bar{F}_{\dot{\alpha}\dot{\beta}})$ the components of the ${\cal N}=2$ vector multiplet, by $(q,\bar{q},\psi_{\alpha},\bar{\psi}_{\dot{\alpha}})$ the components of the ${\cal N}=2$ half-hypermultiplet, and by $\partial_{\alpha\dot{\alpha}}$ the spacetime derivatives.} \label{single letters}
\end{table}
\bea
f^V_{4d}=\frac{-\fp-\fq-\ft+2\fp\fq+\fp\fq/\ft}{(1-\fp)(1-\fq)}\qquad\quad f^{\frac{1}{2}H}_{4d}= \frac{\sqrt{t}-\fp\fq/\sqrt{t}}{(1-\fp)(1-\fq)}.
\eea
This evaluation is analogous to the evaluation of  the chiral multiplet contribution to $2d$ superconformal index by listing its letters, as in table \ref{2dletters}. The multi-particle index of the hypermultiplet is obtained by the plethystic exponent of two copies of $f^{\frac12 H}_{4d}$. Taking the hypermultiplet to be charged under a $U(1)$ symmetry with fugacity $z$,
\be
\II^{H}_{4d}(z;\fp,\fq,\ft)=\prod_{i,j=0}^{\infty}\frac{1-z^\mp \fp^{i+1} \fp^{j+1}/\sqrt{\ft}}{1-z^\pm\sqrt{\ft} \fp^i \fq^j}=: \Gamma(z^\pm\sqrt{\ft};\fp,\fq).
\ee
Here, $\Gamma(z;\fp,\fq)$ is the elliptic gamma function, first defined in \cite{Spiridonov1}. We have used a relatively common notation $\Gamma(z^\pm)=\Gamma(z)\Gamma(z^{-1})$. The elliptic Gamma function enjoys the following remarkable property,
\be\label{gammashift}
\Gamma(z\fp;\fp,\fq)=\theta(z;\fq)\Gamma(z;\fp,\fq),\qquad \Gamma(z\fq;\fp,\fq)=\theta(z;\fp)\Gamma(z;\fp,\fq).
\ee
It is this property that will turn out to be crucial in showing the S-duality invariance of the surface operator index and also in relating our result to that of \cite{Gaiotto:2012xa}.
As in the $2d$ case, the $4d$ vector multiplet contribution changes the measure of the gauge fugacity integral over the Cartan torus. The integral for the $U(N)$ gauge group is,
\be
\II_{4d}=\frac{\kappa^{N}}{N!} \oint \prod_{i=1}^{N}\frac{dz_i}{2\pi i z_i}\frac{\prod_{i,j}\Gamma(\frac{\fp\fq}{\ft} z_i/z_j;\fp,\fq )}{\prod_{i\neq j}\Gamma(z_i/z_j;\fp,\fq)}\ldots,
\ee
where $\kappa=(\fp;\fp)(\fq;\fq)\Gamma(\frac{\fp\fq}{\ft};\fp,\fq)$. The $\ldots$ stand for the contribution from the matter multiplets. This is reminiscent of eq. \eqref{zintegral} for the two dimensional gauge theory integrals.

\subsection{Embedding of $2d$ $(2,2)$ into $4d$ $\CN=2$}\label{symmetries}

We are interested in the half-BPS surface operators in ${\cal N}=2$ gauge theory.  The $4d$ superconformal symmetry group is $SU(2,2|2)$. Its bosonic subgroup is $S[U(2,2)\times U(2)]\sim SU(2,2)\times SU(2)_R\times U(1)_r$. The $SU(2,2)\sim SO(4,2)$ factor is the conformal group in four dimensions and $SU(2)\times U(1)$ is the  $\CN=2$ R symmetry group. We orient the surface operator along the $(x^0,x^1)$ plane. It preserves only $SO(2,2)\times U(1)_{23}\subset SO(4,2)$ part of the conformal group and $U(1)_{\ttL}\times U(1)_{\ttR}\subset SU(2)_{R}\times U(1)_{r}$ part of the R-symmetry group. Here, $SO(2,2)\cong SL(2,\R)_{\ttL}\times SL(2,\R)_{\ttR}$ is the conformal group in two dimensions and $U(1)_{23}$ is the rotation symmetry in the $(x^{2},x^{3})$ plane transverse to the surface operator. Out of the eight supercharges $\QQ_\alpha^{1,2}$ and ${\tilde \QQ}_{\dot \alpha}^{1,2}$ of the four dimensional theory, only $\QQ^2_-,{\tilde \QQ}^1_{\dot -}$ and and $\QQ^1_+,{\tilde \QQ}^2_{\dot +}$  are preserved. The preserved bosonic subgroup $SL(2,\R)_{\ttL}\times U(1)_\ttL$ along with the supersymmetries $\QQ^2_-,{\tilde \QQ}^1_{\dot -}$  generate $SU(1,1|1)_\ttL$. Similarly the remaining charges form $SU(1,1|1)_\ttR$. All in all, a half-BPS surface operator in  ${\cal N}=2$ superconformal theory preserves $SU(1,1|1)_\ttL\times SU(1,1|1)_\ttR \times U(1)_\te$ subgroup of $SU(2,2|2)$. Here $U(1)_\te$ is the commutant of embedding.

Note that we have chosen the orientation of the surface operator so that, among others, it preserves the supercharge $\QQ$ used to define the index. The nontrivial commutation relations of the preserved symmetry algebra are listed in the first column of table \ref{commutation}. Conservation of supercharge $\QQ$ along with all five Cartan generators of $SU(2,2|2)$ makes the index \eqref{indexdef} well defined even in the presence of surface operators. Therefore, we see that even in the presence of a surface operator the superconformal index of the four-dimensional ${\cal N}=2$ gauge theory depends on {\it three} fugacities ($\fp, \fq, \ft$).

\begin{table}
\begin{centering}
\begin{tabular}{|c|c|}
\hline
$4d$ & $2d$\tabularnewline
\hline
\hline
$\{\QQ_{-}^{2},(\QQ_{-}^{2})^{\dagger}\}=(E-h_{01})+(2R+h_{23}-r)$ & $\{\GG_{\ttR}^{+},(\GG_{\ttR}^{+})^{\dagger}\}=2H_{\ttR}+2\nu J_{\ttR}$\tabularnewline
\hline
$\{\widetilde{\QQ}_{\dot{-}}^{1},(\widetilde{\QQ}_{\dot{-}}^{1})^{\dagger}\}=(E-h_{01})-(2R+h_{23}-r)$ & $\{\GG_{\ttR}^{-},(\GG_{\ttR}^{-})^{\dagger}\}=2H_{\ttR}-2\nu J_{\ttR}$\tabularnewline
\hline
$\{\QQ_{+}^{1},(\QQ_{+}^{1})^{\dagger}\}=(E+h_{01})-(2R+h_{23}+r)$ & $\{\GG_{L}^{+},(\GG_{L}^{+})^{\dagger}\}=2H_{\ttL}-2\nupr J_{\ttL}$\tabularnewline
\hline
$\{\widetilde{\QQ}_{\dot{+}}^{2},(\widetilde{\QQ}_{\dot{+}}^{2})^{\dagger}\}=(E+h_{01})+(2R+h_{23}+r)$ & $\{\GG_{\ttL}^{-},(\GG_{\ttL}^{-})^{\dagger}\}=2H_{\ttL}+2\nupr J_{\ttL}$\tabularnewline
\hline
\end{tabular}
\par\end{centering}
\caption{\label{commutation}Important commutation relations of the $SU(1,1|1)\times SU(1,1|1)\times U(1)_{\te}$ subgroup of the $4d$ ${\cal N}=2$ superconformal algebra and $2d$ $(2,2)$ superconformal algebra are listed in the first and second column respectively. This leads to a map between $4d$ charges and $2d$ charges as described in the text.}
\end{table}

In order to get the exact map between the four dimensional quantum numbers and the two dimensional ones, we compare the above algebra with the $(2_\nu,2_\nupr)$ algebra in two dimensions, where the subscripts $\nu$ and $\nupr$ are spectral flow parameters. Let us denote the supercharges in the left moving and right moving sectors as $\GG_\ttL^\pm$ and $\GG_\ttR^\pm$ respectively. The scaling and R-symmetry generators are denoted as $H_{\ttL,\ttR}$ and $J_{\ttL,\ttR}$. With these notations, the nontrivial commutation relations of the $2d$ $(2,2)$ superconformal algebra are as shown in the second column of table \ref{commutation}. Comparing the first and second column leads to the following identification of the supercharges:
\be
\GG_\ttR^+=\QQ^2_-,\quad \GG_\ttR^-={\tilde \QQ}^1_{\dot -},\quad \GG_\ttL^+=\QQ^1_+,\quad \GG_\ttL^-={\tilde \QQ}^2_{\dot +}
\ee
and bosonic charges:
\bea \label{2d4dmap}
H\equiv H_\ttL+H_\ttR=E, &&s\equiv H_\ttL-H_\ttR=h_{01}\nonumber\\
J_A\equiv J_\ttL-J_\ttR=2r, &&J_V\equiv J_\ttL+J_\ttR=4R+2h_{23}
\eea
along with $\nu=\nupr=1/2$ indicating the relevant $(2,2)$ superconformal algebra is the one in the NSNS sector. This justifies our  choice of the $2d$ index as the superconformal index in the NSNS sector. In addition to \eqref{2d4dmap}, the charge $R+h_{23}$ generating $U(1)_\te$ becomes the flavor symmetry of the embedded $(2,2)$ algebra. Comparing \eqref{2dindexdef2} and \eqref{indexdef}, this map leads to the identification of the $4d$ fugacities $(\fp,\fq,\ft)$ and $2d$ fugacities $(q,t,\te)$:
\be
q=\fq, \qquad t=\fp\fq/\ft, \qquad \te=\fp^2/\ft.
\ee
Here, $\te$ is the fugacity of the $2d$ theory living on the surface operator that couples to $U(1)_\te$ flavor symmetry. Now we are in the position to compute the index of the surface operator as $2d$-$4d$ coupled system.

\subsection{Duality check}\label{dualitycheck}

\begin{figure}[t]
\centering
\includegraphics[scale=0.45]{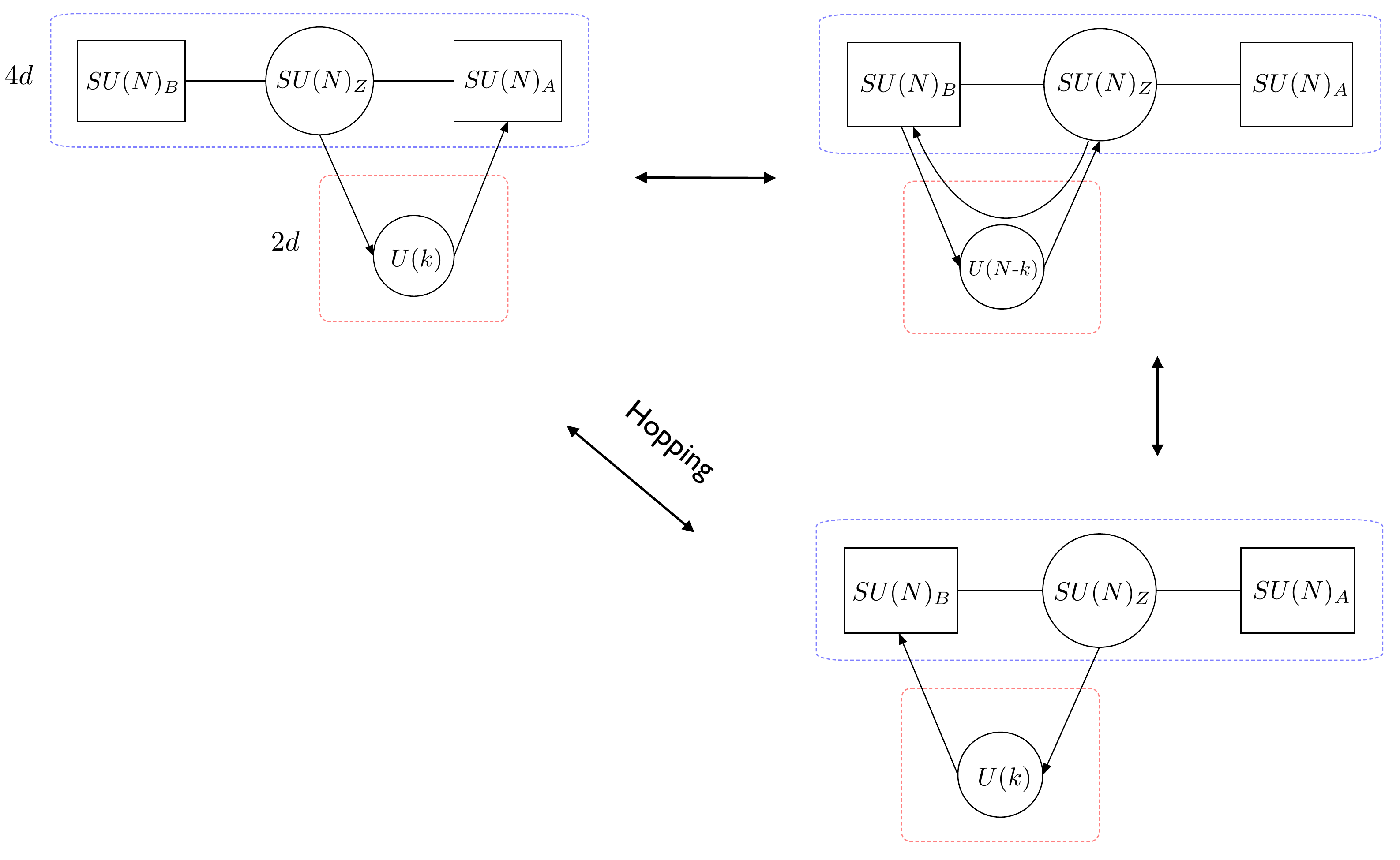}
\caption{The first figure describes the surface operator in a duality frame where $\TS$ is a U(k) gauge theory and couples to nodes  $Z$ and $A$. This configuration is dual to the $U(N-k)$ gauge theory coupled to nodes $Z$ and $B$. The curved arrow denotes the $2d$ meson field. The $2d$ Seiberg duality on the $U(N-k)$ node give back the $U(k)$ gauge theory. The two steps together constitute the hopping of the surface operator. }
\label{hopping}
\end{figure}

The index of the $SU(N)$ $\CN=2$ SCQCD in the absence of surface operator is:
\be
\II_{4d}=\frac{\kappa^{N-1}}{N!} \oint_{{\mathbb T}^{N-1}} \prod_{i=1}^{N-1}\frac{dz_i}{2\pi i z_i}\frac{\prod_{i,j}\Gamma(\frac{\fp\fq}{\ft} z_i/z_j )}{\prod_{i\neq j}\Gamma(z_i/z_j)} \prod_{i,j}\Gamma((x z_i a_j)^{\pm}\sqrt{\ft})\Gamma((y\frac{b_i}{z_j})^{\pm}\sqrt{\ft}).
\ee
We have suppressed the parameters $(;\fp,\fq)$ of the elliptic Gamma function. The variables $a_i$ and $b_i$ are the fugacities for $SU(N)_A$ and $SU(N)_B$ flavor symmetries, $x$ and $y$ are fugacities for the two remaining $U(1)$ symmetries.
Coupling to $2d$ theory, as shown in figure \ref{N2general} for $n=2$ and $p_1=k$, introduces additional ``matter" from the point of view of the $4d$ theory. In fact, the $2d$ theory is exactly the one studied in the context of the Seiberg-type duality version 2 in section \ref{nonabelian-without}. We borrow its index $\II^{(2)}_k({\bf z},{\bf a},c;q,t)$ and multiply it to the integrand. We have to take the chiral fields of the $2d$ theory to be charged under the $U(1)_e$ symmetry as well as the $U(1)_x$ symmetry of the four dimensional theory. The precise charge assignment is achieved by setting $c^2=xe^{-1/2}$.
Importantly, we need to make the change of variables as outlined in the section \ref{symmetries} i.e. $q=\fq, \,t=\fp\fq/\ft, \,c^2=x\sqrt{\ft}/\fp$. This gives,
\be
\II_{2d\mbox{-}4d}&=&\frac{\kappa^{N-1}}{N!} \oint_{{\mathbb T}^{N-1}} \prod_{i=1}^{N-1}\frac{dz_i}{2\pi i z_i}\frac{\prod_{i,j}\Gamma(\frac{\fp\fq}{\ft} z_i/z_j )}{\prod_{i\neq j}\Gamma(z_i/z_j)} \prod_{i,j}\Gamma((x z_i a_j)^{\pm}\sqrt{\ft})\Gamma((y\frac{b_i}{z_j})^{\pm}\sqrt{\ft})\nonumber\\
&\times &\sum_{\{i_\alpha\}} \prod_{s\in \{i_\alpha\}}  \prod_j \Delta(\frac{x\sqrt{\ft}}{\fp}a_j z_s;\fq,\frac{\fp\fq}{\ft})\prod_{s\in \{i_\alpha\}}  \prod_{r\in \bar{\{i_{\alpha}\}}}\Delta(z_r/z_s;\fq,\frac{\fp\fq}{\ft}).
\ee
Note that due to ${\bf z}$ dependence of the $2d$ index, it can't be separated from the integral. This is the only signature of the $2d$-$4d$ coupling in the computation of superconformal index. We can simplify the integrand using the property eq. \eqref{gammashift}. Absorbing the first factor in the second line into $\Gamma((x z_i a_j)^{\pm}\sqrt{\ft})$ we get,
\be
\II_{2d\mbox{-}4d}&=&\frac{\kappa^{N-1}}{N!} \oint_{{\mathbb T}^{N-1}} \prod_{i=1}^{N-1}\frac{dz_i}{2\pi i z_i}\frac{\prod_{i,j}\Gamma(\frac{\fp\fq}{\ft} z_i/z_j )}{\prod_{i\neq j}\Gamma(z_i/z_j)} \prod_{i,j}\Gamma((y\frac{b_i}{z_j})^{\pm}\sqrt{\ft})\nonumber\\
&\times &\sum_{\{i_\alpha\}} \prod_{s\in \{i_\alpha\}}  \prod_{r\in \bar{\{i_{\alpha}\}}}\Delta(z_r/z_s;\fq,\frac{\fp\fq}{\ft}) \prod_{j}\Gamma((p^{-1}x z_s a_j)^{\pm}\sqrt{\ft})\Gamma((x z_r a_j)^{\pm}\sqrt{\ft}) .
\ee
For $s\in \{i_\alpha\}$ and $r\in \bar{\{i_\alpha\}}$, we make the change of the dummy variable $z_s\to{\tilde z}_s=p^{-1}z_s$ and $z_r\to {\tilde z}_r=z_r$. After playing with it a bit we get a very elegant result,
\be
\II_{2d\mbox{-}4d}&=&\frac{\kappa^{N-1}}{N!} \oint_{{\mathbb T}^{N-1}} \prod_{i=1}^{N-1}\frac{d{\tilde z}_i}{2\pi i {\tilde z}_i}\frac{\prod_{i,j}\Gamma(\frac{\fp\fq}{\ft} {\tilde z}_i/{\tilde z}_j )}{\prod_{i\neq j}\Gamma({\tilde z}_i/{\tilde z}_j)} \prod_{i,j}\Gamma((x {\tilde z}_i a_j)^{\pm}\sqrt{\ft})\Gamma((y\frac{b_i}{{\tilde z}_j})^{\pm}\sqrt{\ft})\\
&\times & \prod_{i,j} \Delta(\frac{y\sqrt{\ft}}{\fp}\frac{b_j}{ {\tilde z}_i};\fq,\frac{\fp\fq}{\ft})\sum_{\bar{\{i_\alpha\}}} \prod_{r\in \bar{\{i_\alpha\}}}  \prod_j \Delta(\frac{\sqrt{\ft}}{y}\frac{z_r}{ {b}_j};\fq,\frac{\fp\fq}{\ft})\prod_{s\in \{i_\alpha\}}  \prod_{r\in \bar{\{i_{\alpha}\}}}\Delta({\tilde z}_s/{\tilde z}_r;\fq,\frac{\fp\fq}{\ft})\nonumber.
\ee
We see that in this description the contribution of the $2d$ part is:
\be
\II_{2d}=\Big(\prod_{i,j}\Delta(\frac{\ft \fp^{-1}}{{\tilde c}^2}\frac{1}{{\tilde z}_i {\tilde b}_j};\fq,\frac{\fp\fq}{\ft})\Big) \,\,\II_{N-k}^{(2)}({\tilde{\bf z}},{\tilde{\bf b}},{\tilde c};\fq,\frac{\fp\fq}{\ft}).
\ee
where ${\tilde b}_i=1/b_i$ and ${\tilde c}^2=\sqrt{\ft}/y=\frac{qe^{-1/2}}{ty}$. This is precisely the index of the Seiberg (or Hori-Tong) dual of the $2d$ theory that we started off with. Moreover, while the original $\TS$ coupled to the $SU(N)_Z$ and $SU(N)_A$ nodes of the $4d$ theory, the new theory couples to $SU(N)_Z$ and $SU(N)_B$ nodes! Now we can perform the $2d$ Seiberg duality to go back to the $U(k)$ gauge theory. All in all, our surface operator has ``hopped" one node of the quiver as shown in figure \ref{hopping}. The hopping readily generalizes to longer quivers.

\begin{figure}[h]
\centering
\includegraphics[scale=0.45]{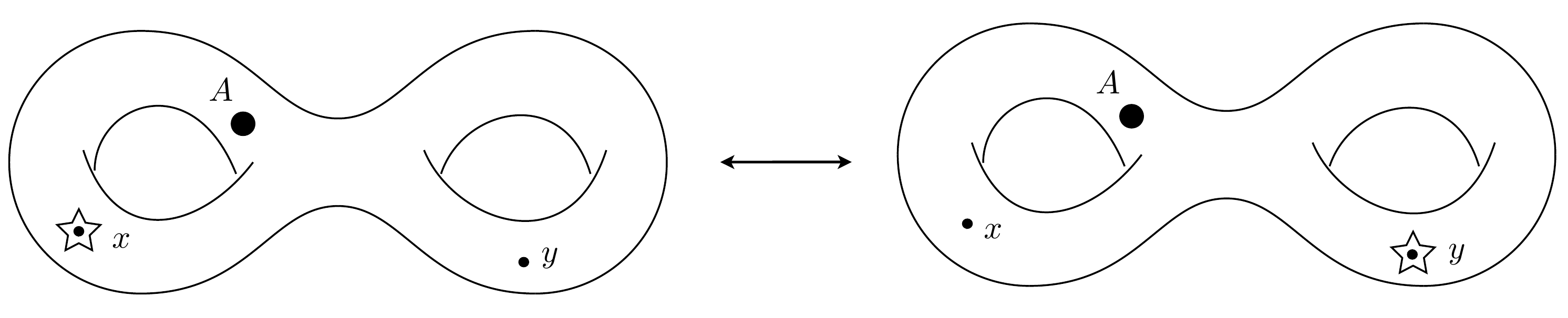}
\caption{The small dots labeled by $x$ and $y$ stand for punctures corresponding to $U(1)_x$ and $U(1)_y$ flavor symmetries respectively. The large dot labeled by $A$ denotes the puncture with flavor symmetry $SU(N)_A$. The star decoration indicates coupling of  the surface operator to the $U(1)$ puncture.}
\label{dualityinvariance}
\end{figure}

A large class $\CS$ of $\CN=2$ superconformal theories is obtained by compactifying M5 branes on Riemann surface. The surface operator in such theories is parametrized by a marked point on the Riemann surface. 
Depending on which minimal puncture (puncture with $U(1)$ flavor symmetry) the marked point approaches,  we get different dual weakly coupled descriptions. What we have shown is the that the superconformal index is same in all such duality frames. This implies that our construction of the surface operator is invariant under the generalized S-duality. 
It is important to comment that although we have computed the surface operator index in gauge theory, the result about the S-duality invariance of the index holds for general theories of class $\CS$. This is illustrated with a simple example in figure \ref{dualityinvariance}: Take the degeneration limit at the minimal puncture $x$ i.e. (almost) decouple the associated hypermultiplet $\oplus$ surface operator system. Use S-dualities in the rest of the theory to move the puncture $y$ next to $x$. The surface operator can now hop to $y$. After the hopping take the degeneration limit at $y$ and use S-dualities to go back to the original duality frame. 

\section{Surface operators from vortex strings}\label{vortex}

Yet another way of thinking about the surface operators is as infinite tension, zero thickness scaling limit of the ``semi-local" vortex string. Although this approach doesn't produce all the surface operators, it does provide us with a distinguished class whose $\TS$ can be readily identified.

In this approach, in order to construct the surface operator in $\cT$, we have to consider vortex solutions in $SU(N)\times U(N)'$ quiver gauge theory. 
\begin{center}
\includegraphics[scale=0.25]{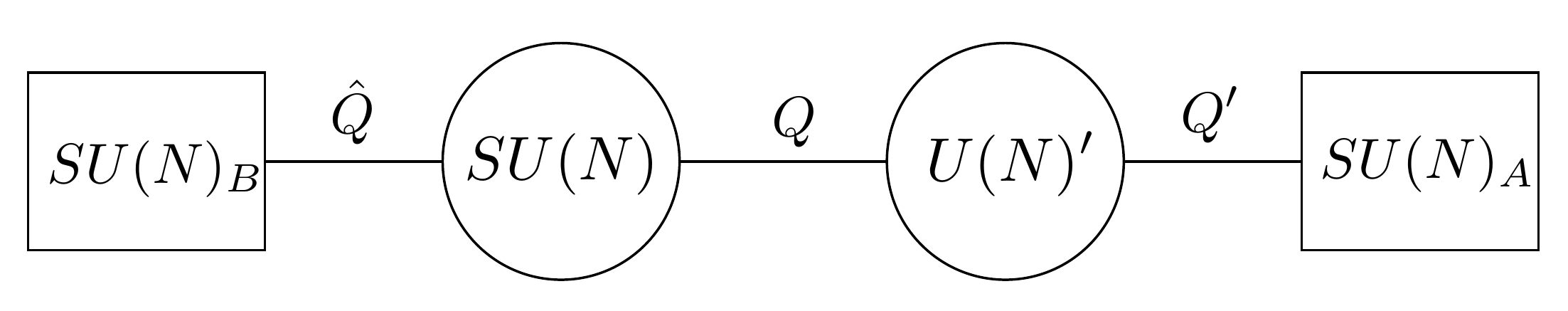}
\end{center}
Let $\xi$ be the FI parameter for the $U(N)'$ gauge group. In the phase of the theory where $U(N)'$ is Higgsed, the fields ${Q^\prime}_s^k$ transforming in the bi-fundamental representation of $U(N)'\times SU(N)_A$ get a non-zero vev $\sqrt{\xi}$. Dynamical vortex configurations arise in this phase when this vev  gets angular dependence in $(x^2,x^3)$ plane. Schematically, the profile of $Q$ and $Q^\prime$ in $(x^2,x^3)\sim(r,\theta)$ plane goes as,
\be
Q^\prime\sim\sqrt{\xi}\frac{re^{in\theta}}{\sqrt{r^2+\rho^2}},\qquad Q\sim\sqrt{\xi}\frac{\rho}{\sqrt{r^2+\rho^2}}.
\ee
The parameter $\rho$ is interpreted as the thickness of the vortex string. The tension of the vortex is $2\pi \xi$. The gauge group $U(N)'$ is completely Higgsed and  is ``locked" with the flavor group $SU(N)_A$. In order to obtain the surface operator in the Higgsed theory, we have to take the limit
\be\label{scalinglimit}
\rho\to0,\qquad  \xi \to \infty,\qquad \qquad \rho \sqrt{\xi}=\mbox{const.}
\ee
This completely removes the fields $Q^\prime$ and leads to the reduced theory: $SU(N)$ SCQCD but with singularity for the $Q$ field. This is exactly the singularity resulting from \eqref{N2surface}. The $SU(N)$ gauge field also gets the singularity as expected from \eqref{N2surface}.

\begin{figure}[h]
\centering
\includegraphics[scale=0.29]{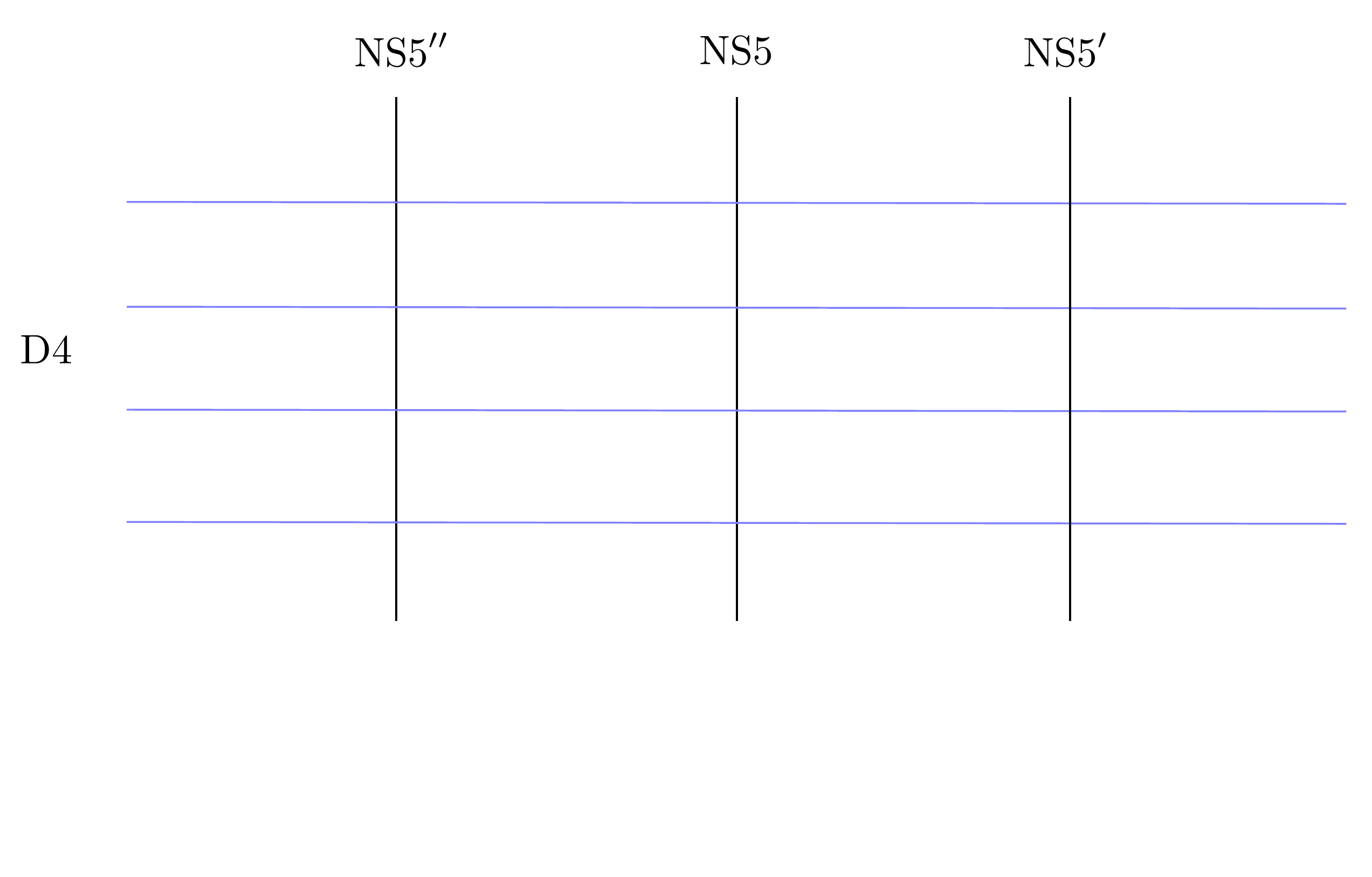}
\includegraphics[scale=0.3]{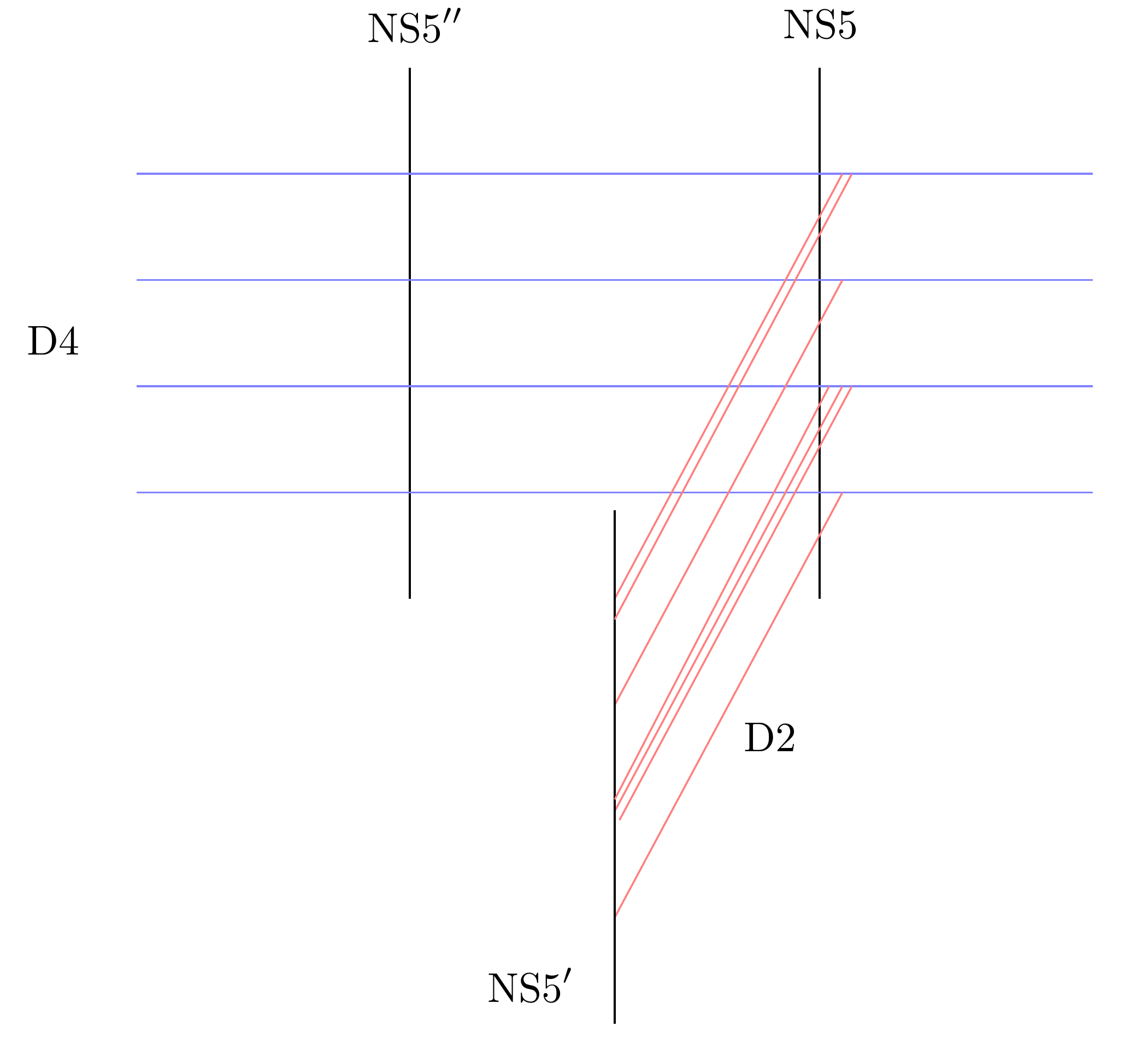}
\caption{The first figure shows the Hanany-Witten brane setup for the ${\cal N}=2$ $SU(N)\times SU(N)$ quiver gauge theory. As we move NS$'$ brane away from D4 branes, new D2 branes develop between NS5$'$ and D4. They represent half-BPS vortex string solutions in the $\CN=2$ theory. Their infinite tension limit engineers the surface operators in the \emph{reduced} theory i.e. SCQCD. This limit corresponds to moving the NS5$'$ brane off to infinity.}
\label{N2vortex}
\end{figure}

The quickest way to summarize the discussion is through the brane diagram \ref{N2vortex}. Higgsing of $U(N)'$ corresponds to moving the NS5$^\prime$ brane away from the D4 branes. Vortices in this phase map to D2 branes stretching between NS5$^\prime$ and D4 branes. The scaling limit \eqref{scalinglimit} means taking the NS5$^\prime$ brane all the way to infinity, corresponding to the surface operator in SCQCD.

The vortex string approach to surface operators was used in \cite{Gaiotto:2012xa} to compute its index. We will see that the superconformal index computed in a more direct way matches with the index computed there.

\subsection{Their index}\label{surfaceindex}
\begin{figure}[h]
\centering
\includegraphics[scale=0.4]{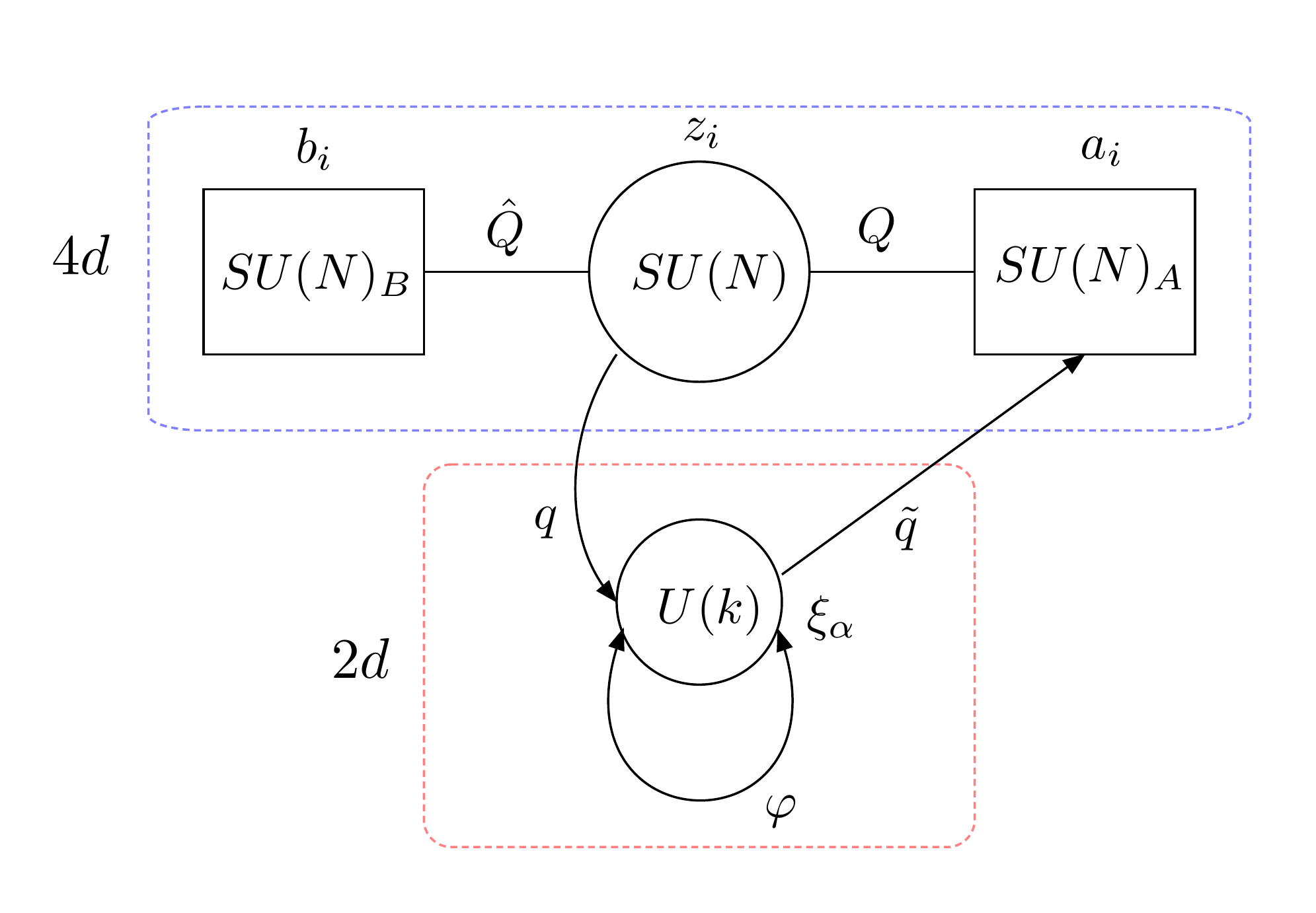}
\caption{The $2d$-$4d$ quiver gauge theory corresponding to the brane setup of figure \ref{N2vortex}. This is very similar to the system studied in section \ref{dualitycheck}. The only difference is that the $2d$ theory has an additional adjoint chiral field $\varphi$. The fugacities corresponding to various gauge and symmetries are also indicated in the figure.}
\label{N2vortexquiver}
\end{figure}

\begin{figure}[t]
\centering
\includegraphics[scale=0.40]{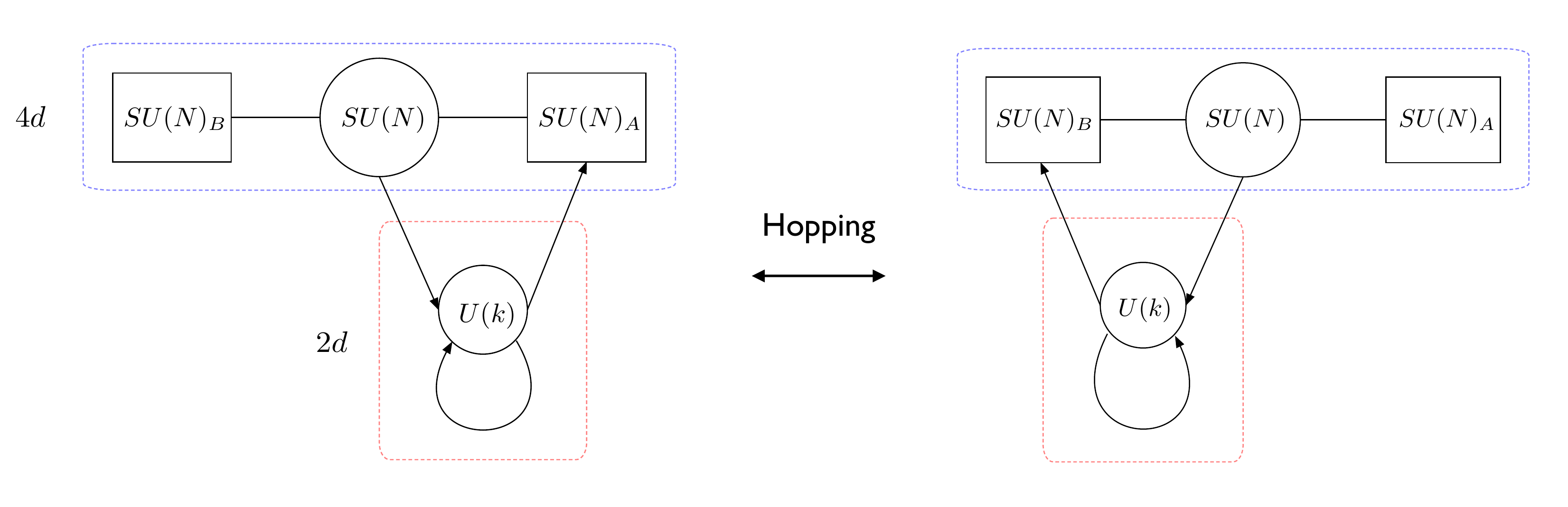}
\caption{Hopping of the surface operator coming from vortex strings}
\label{hopping2}
\end{figure}

The gauge theory system corresponding to the brane setup in  figure \ref{N2vortex} can be easily read off. Its quiver diagram is given in figure \ref{N2vortexquiver}. This $2d$-$4d$ system is very similar to the one whose index was computed in the last section. The difference being the additional $2d$ adjoint chiral multiplet in this case. The index of the $2d$ part was computed in section \ref{nonabelian-with}. Borrowing that result, the index of this coupled system is:

\be
\II_{2d\mbox{-}4d}^{\rm vortex}&=&\frac{\kappa^{N-1}}{N!} \oint_{{\mathbb T}^{N-1}} \prod_{i=1}^{N-1}\frac{dz_i}{2\pi i z_i}\frac{\prod_{i,j}\Gamma(\frac{\fp\fq}{\ft} z_i/z_j )}{\prod_{i\neq j}\Gamma(z_i/z_j)} \prod_{i,j}\Gamma((x z_i a_j)^{\pm}\sqrt{\ft})\Gamma((y\frac{b_i}{z_j})^{\pm}\sqrt{\ft})\nonumber\\
&\times &\II^{(4)}_k({\bf z},{\bf a},(x\sqrt{\ft}/\fp)^{1/2},\fp^{-1};\fq,\frac{\fp\fq}{\ft}).
\ee
The explicit expression for $\II^{(4)}_k$ is found in eq. \eqref{withadj}. Here, the adjoint chiral multiplet of the $2d$ theory is also charged under the $U(1)_e$ symmetry, the commutant of $(2,2)$ algebra inside $4d$ $\CN=2$ algebra. Its charge can be encoded in the index by setting $d=(t/q)e^{-1}=\fp^{-1}$.
Like in the previous case, using \eqref{gammashift}, the factors of $\Delta$ function can be absorbed into $\Gamma$ function.
\be
\II_{2d\mbox{-}4d}^{\rm vortex}&=&\frac{\kappa^{N-1}}{N!} \oint_{{\mathbb T}^{N-1}} \prod_{i=1}^{N-1}\frac{dz_i}{2\pi i z_i}\frac{\prod_{i,j}\Gamma(\frac{\fp\fq}{\ft} z_i/z_j )}{\prod_{i\neq j}\Gamma(z_i/z_j)} \prod_{i,j}\Gamma((y\frac{b_i}{z_j})^{\pm}\sqrt{\ft})\nonumber\\
&\times & \sum_{\{n_i\}} \prod_{i,j}^{N} \Gamma((\fp^{-n_i}x z_i a_j)^{\pm}\sqrt{\ft})
\prod_{n=0}^{n_{i}-1}\Delta(\fp^{n-n_{j}}z_{j}/z_{i}).
\ee
The $2d$ contribution can be conveniently summarized in terms of a difference operator $\CS_k$ as follows:
\be
\II_{2d\mbox{-}4d}^{\rm vortex}&=&\frac{\kappa^{N-1}}{N!} \oint_{{\mathbb T}^{N-1}} \prod_{i=1}^{N-1}\frac{dz_i}{2\pi i z_i}\frac{\prod_{i,j}\Gamma(\frac{\fp\fq}{\ft} z_i/z_j )}{\prod_{i\neq j}\Gamma(z_i/z_j)} \Big(\CS_k\cdot\prod_{i,j}\Gamma((x z_i a_j)^{\pm}\sqrt{\ft})\Big)\prod_{i,j}\Gamma((y\frac{b_i}{z_j})^{\pm}\sqrt{\ft}).\nonumber
\ee
where,
\be
\CS_k\cdot f(z_i)=\sum_{\{n_i\}} \prod_{i,j}^{N} f(\fp^{-n_i} z_i)
\prod_{n=0}^{n_{i}-1}\Delta(\fp^{n-n_{j}}z_{j}/z_{i}).
\ee
This is exactly the difference operator studied in \cite{Gaiotto:2012xa} (modulo an overall fractional shift by $\fp^{k/N}$). This difference operator was conjectured to introduce a half-BPS surface operator in an $\CN=2$ superconformal field theory. Here we have explicitly verified the conjecture for quiver type theories. Thanks to the ``self-adjointness" of $\CS_k$ proved in \cite{Gaiotto:2012xa}, we have:
\be
\II_{2d\mbox{-}4d}^{\rm vortex}&=&\frac{\kappa^{N-1}}{N!} \oint_{{\mathbb T}^{N-1}} \prod_{i=1}^{N-1}\frac{dz_i}{2\pi i z_i}\frac{\prod_{i,j}\Gamma(\frac{\fp\fq}{\ft} z_i/z_j )}{\prod_{i\neq j}\Gamma(z_i/z_j)} \prod_{i,j}\Gamma((x z_i a_j)^{\pm}\sqrt{\ft})\Big(\CS_k\cdot\prod_{i,j}\Gamma((y\frac{b_i}{z_j})^{\pm}\sqrt{\ft})\Big).\nonumber
\ee
This means, just like the surface operator studied in \ref{dualitycheck}, the index of this surface operator is same after hopping along the quiver, see figure \ref{hopping2}. Repeating the arguments at the end of section \ref{dualitycheck} we conclude that the index of this type of surface operator is also invariant under generalized S-duality.

\section*{Acknowledgements}
The authors would like to thank Yu Nakayama, Hirosi Ooguri, Pavel Putrov and Shlomo Razamat for interesting discussions. Authors are especially grateful to Anton Kapustin for his valuable comments.
The work of A.G. is supported in part by the John A. McCone fellowship and by DOE Grant DE-FG02-92- ER40701. The work of S.G. is supported in part by DOE Grant DE-FG03-92-ER40701FG-02 and in part by NSF Grant PHY-0757647. Opinions and conclusions expressed here are those of the authors and do not necessarily reflect the views of funding agencies.

\appendix

\section{Properties of the multiplet index}\label{properties}

\subsection{Chiral multiplet index}
It is useful to note the following properties of the $\theta(z;q):=\prod_{i=0}^\infty(1-x q^i)(1-q^{i+1}/z)$ function.
\be
\theta(qz;q)= -\frac{1}{z}\theta(z;q),\qquad \theta(z^{-1};q) =  \theta(qz;q)=-\frac{1}{z}\theta(z;q)
\ee
They translate into the following properties of the chiral multiplet index $\Delta(z;q,t)$:
\be\label{Deltashift}
&&\Delta(zq;q,t)  = \frac{1}{t}\Delta(z;q,t) ,\qquad \qquad \quad\, \Delta(z;q,qt) = -\frac{1}{zt}\Delta(z;q,t), \label{qshift}\\
&&\Delta(zq/t;q,t) = 1/\Delta(z^{-1};q,t),\qquad \quad\Delta(z^{-1};q,t^{-1})=\frac{1}{t}\Delta(z;q,t). \label{Deltainverse}
\ee
If we define the shift operators ${p}_z$ and ${p}_t$ satisfying the $q$-commutations: ${p}_x {x}=q x {p}_x $, then \eqref{Deltashift} can be cast into an operator equations satisfied by $\Delta(z;q,t)$,
\be
{p}_z -1/{t}=0,\qquad p_t+1/zt=0.
\ee

\subsection*{Modular property}
The $\theta(z;q)$ has a modular property,
\begin{eqnarray}
\theta(e^{2\pi i(-\frac{\xi}{\tau})};e^{2\pi i(-\frac{1}{\tau})}) & = & e^{i\pi B(\xi,\tau)}\theta(e^{2\pi i\xi};e^{2\pi i\tau})\nonumber \\
B(\xi,\tau) & = & \frac{\xi^{2}}{\tau}-\xi-\frac{\xi}{\tau}+\frac{1}{6}(\tau+\frac{1}{\tau})+\frac{1}{2}
\end{eqnarray}
The index of the chiral multiplet is simply the ratio of two $\theta$ functions, so it inherits the modular property  as expected from any $2d$ superconformal index.
\be
\Delta(e^{2\pi i(-\frac{\xi}{\tau})};e^{2\pi i(-\frac{1}{\tau})},e^{2\pi i(-\frac{\sigma}{\tau})}) & = & \frac{e^{i\pi B(\xi+\sigma,\tau)}\theta(e^{2\pi i(\xi+\sigma)};e^{2\pi i\tau})}{e^{i\pi B(\xi,\tau)}\theta(e^{-2\pi i\xi};e^{2\pi i\tau})}\nonumber \\
 & = & e^{i\pi\frac\sigma\tau(2\xi+\sigma-\tau-1)}\Delta(e^{2\pi i\xi};e^{2\pi i\tau},e^{2\pi i\sigma})\label{Deltamodular}
\ee
The modular property \eqref{Deltamodular} makes the shift symmetries \eqref{Deltashift} manifest.

\subsection{Vector multiplet index}
The vector multiplet $\II_V(q.t)=(q;q)^2/\theta(t,q)$ satisfies the properties:
\be \label{vectorinverse}\label{vectorshift}
\II_V(q,qt)= \II_{V}(t^{-1},q)=-t\,\,\II_{V}(t,q) \qquad \Rightarrow \qquad  p_t +t=0.
\ee
It has interesting modular transformation properties as well. Making use of \eqref{vectorresidue},
\begin{eqnarray}
\frac{\II_{V}(e^{2\pi i(-\frac{\sigma}{\tau})},e^{2\pi i(-\frac{1}{\tau})})}{\II_{V}(e^{2\pi i\sigma},e^{2\pi i\tau})} & = & \lim_{\xi\to0}\frac{\Delta(e^{2\pi i\xi};e^{2\pi i\tau},e^{2\pi i\sigma})}{\Delta(e^{2\pi i(-\frac{\xi}{\tau})};e^{2\pi i(-\frac{1}{\tau})},e^{2\pi i(-\frac{\sigma}{\tau})})}\frac{1-e^{2\pi i\xi}}{1-e^{2\pi i(-\frac{\xi}{\tau})}}\nonumber \\
\II_{V}(e^{2\pi i(-\frac{\sigma}{\tau})},e^{2\pi i(-\frac{1}{\tau})}) & = & -\tau e^{-i\pi\frac{\sigma}{\tau}(\sigma-\tau-1 )}\II_{V}(e^{2\pi i\sigma},e^{2\pi i\tau}).
\end{eqnarray}

\bibliographystyle{JHEP}
\bibliography{SurfaceIndex}

\end{document}